\documentclass[reprint,amsmath,amssymb, aps,prd, floatfix ]{revtex4-2}

\usepackage{graphicx}
\usepackage{dcolumn}
\usepackage{bm}
\usepackage{xcolor}
\usepackage{siunitx}
\usepackage{multirow}
\usepackage[caption=false]{subfig}
\usepackage{ragged2e}
\DeclareCaptionJustification{justified}{\justifying}
\usepackage[multiple]{footmisc}
\usepackage[bookmarks=false]{hyperref}

\begin{document}

\title{Measurements of the $\nu_{\mu}$ and $\bar{\nu}_{\mu}$-induced Coherent Charged Pion Production Cross Sections on $^{12}\textup{C}$ by the T2K Experiment.}


\newcommand{\INSTHD}{\affiliation{University Autonoma Madrid, Department of Theoretical Physics, 28049 Madrid, Spain}}
\newcommand{\INSTEE}{\affiliation{University of Bern, Albert Einstein Center for Fundamental Physics, Laboratory for High Energy Physics (LHEP), Bern, Switzerland}}
\newcommand{\INSTFE}{\affiliation{Boston University, Department of Physics, Boston, Massachusetts, U.S.A.}}
\newcommand{\INSTGA}{\affiliation{University of California, Irvine, Department of Physics and Astronomy, Irvine, California, U.S.A.}}
\newcommand{\INSTI}{\affiliation{IRFU, CEA, Universit\'e Paris-Saclay, F-91191 Gif-sur-Yvette, France}}
\newcommand{\INSTGB}{\affiliation{University of Colorado at Boulder, Department of Physics, Boulder, Colorado, U.S.A.}}
\newcommand{\INSTFG}{\affiliation{Colorado State University, Department of Physics, Fort Collins, Colorado, U.S.A.}}
\newcommand{\INSTFH}{\affiliation{Duke University, Department of Physics, Durham, North Carolina, U.S.A.}}
\newcommand{\INSTJA}{\affiliation{E\"{o}tv\"{o}s Lor\'{a}nd University, Department of Atomic Physics, Budapest, Hungary}}
\newcommand{\INSTEF}{\affiliation{ETH Zurich, Institute for Particle Physics and Astrophysics, Zurich, Switzerland}}
\newcommand{\INSTIE}{\affiliation{CERN European Organization for Nuclear Research, CH-1211 Gen\'eve 23, Switzerland}}
\newcommand{\INSTEG}{\affiliation{University of Geneva, Section de Physique, DPNC, Geneva, Switzerland}}
\newcommand{\INSTHJ}{\affiliation{University of Glasgow, School of Physics and Astronomy, Glasgow, United Kingdom}}
\newcommand{\INSTDG}{\affiliation{H. Niewodniczanski Institute of Nuclear Physics PAN, Cracow, Poland}}
\newcommand{\INSTCB}{\affiliation{High Energy Accelerator Research Organization (KEK), Tsukuba, Ibaraki, Japan}}
\newcommand{\INSTIB}{\affiliation{University of Houston, Department of Physics, Houston, Texas, U.S.A.}}
\newcommand{\INSTED}{\affiliation{Institut de Fisica d'Altes Energies (IFAE) - The Barcelona Institute of Science and Technology, Campus UAB, Bellaterra (Barcelona) Spain}}
\newcommand{\INSTJC}{\affiliation{Institut f\"ur Physik, Johannes Gutenberg-Universit\"at Mainz, Staudingerweg 7, 55128 Mainz, Germany}}
\newcommand{\INSTEC}{\affiliation{IFIC (CSIC \& University of Valencia), Valencia, Spain}}
\newcommand{\INSTHH}{\affiliation{Institute For Interdisciplinary Research in Science and Education (IFIRSE), ICISE, Quy Nhon, Vietnam}}
\newcommand{\INSTEI}{\affiliation{Imperial College London, Department of Physics, London, United Kingdom}}
\newcommand{\INSTGF}{\affiliation{INFN Sezione di Bari and Universit\`a e Politecnico di Bari, Dipartimento Interuniversitario di Fisica, Bari, Italy}}
\newcommand{\INSTBE}{\affiliation{INFN Sezione di Napoli and Universit\`a di Napoli, Dipartimento di Fisica, Napoli, Italy}}
\newcommand{\INSTBF}{\affiliation{INFN Sezione di Padova and Universit\`a di Padova, Dipartimento di Fisica, Padova, Italy}}
\newcommand{\INSTBD}{\affiliation{INFN Sezione di Roma and Universit\`a di Roma ``La Sapienza'', Roma, Italy}}
\newcommand{\INSTEB}{\affiliation{Institute for Nuclear Research of the Russian Academy of Sciences, Moscow, Russia}}
\newcommand{\INSTHI}{\affiliation{International Centre of Physics, Institute of Physics (IOP), Vietnam Academy of Science and Technology (VAST), 10 Dao Tan, Ba Dinh, Hanoi, Vietnam}}
\newcommand{\INSTJD}{\affiliation{ILANCE, CNRS – University of Tokyo International Research Laboratory, Kashiwa, Chiba 277-8582, Japan}}
\newcommand{\INSTHA}{\affiliation{Kavli Institute for the Physics and Mathematics of the Universe (WPI), The University of Tokyo Institutes for Advanced Study, University of Tokyo, Kashiwa, Chiba, Japan}}
\newcommand{\INSTID}{\affiliation{Keio University, Department of Physics, Kanagawa, Japan}}
\newcommand{\INSTIF}{\affiliation{King's College London, Department of Physics, Strand, London WC2R 2LS, United Kingdom}}
\newcommand{\INSTCC}{\affiliation{Kobe University, Kobe, Japan}}
\newcommand{\INSTCD}{\affiliation{Kyoto University, Department of Physics, Kyoto, Japan}}
\newcommand{\INSTEJ}{\affiliation{Lancaster University, Physics Department, Lancaster, United Kingdom}}
\newcommand{\INSTII}{\affiliation{Lawrence Berkeley National Laboratory, Berkeley, CA 94720, USA}}
\newcommand{\INSTBA}{\affiliation{Ecole Polytechnique, IN2P3-CNRS, Laboratoire Leprince-Ringuet, Palaiseau, France}}
\newcommand{\INSTFC}{\affiliation{University of Liverpool, Department of Physics, Liverpool, United Kingdom}}
\newcommand{\INSTFI}{\affiliation{Louisiana State University, Department of Physics and Astronomy, Baton Rouge, Louisiana, U.S.A.}}
\newcommand{\INSTIH}{\affiliation{Joint Institute for Nuclear Research, Dubna, Moscow Region, Russia}}
\newcommand{\INSTHB}{\affiliation{Michigan State University, Department of Physics and Astronomy,  East Lansing, Michigan, U.S.A.}}
\newcommand{\INSTCE}{\affiliation{Miyagi University of Education, Department of Physics, Sendai, Japan}}
\newcommand{\INSTDF}{\affiliation{National Centre for Nuclear Research, Warsaw, Poland}}
\newcommand{\INSTFJ}{\affiliation{State University of New York at Stony Brook, Department of Physics and Astronomy, Stony Brook, New York, U.S.A.}}
\newcommand{\INSTGJ}{\affiliation{Okayama University, Department of Physics, Okayama, Japan}}
\newcommand{\INSTCF}{\affiliation{Osaka Metropolitan University, Department of Physics, Osaka, Japan}}
\newcommand{\INSTGG}{\affiliation{Oxford University, Department of Physics, Oxford, United Kingdom}}
\newcommand{\INSTIC}{\affiliation{University of Pennsylvania, Department of Physics and Astronomy,  Philadelphia, PA, 19104, USA.}}
\newcommand{\INSTGC}{\affiliation{University of Pittsburgh, Department of Physics and Astronomy, Pittsburgh, Pennsylvania, U.S.A.}}
\newcommand{\INSTFA}{\affiliation{Queen Mary University of London, School of Physics and Astronomy, London, United Kingdom}}
\newcommand{\INSTE}{\affiliation{University of Regina, Department of Physics, Regina, Saskatchewan, Canada}}
\newcommand{\INSTGD}{\affiliation{University of Rochester, Department of Physics and Astronomy, Rochester, New York, U.S.A.}}
\newcommand{\INSTHC}{\affiliation{Royal Holloway University of London, Department of Physics, Egham, Surrey, United Kingdom}}
\newcommand{\INSTBC}{\affiliation{RWTH Aachen University, III. Physikalisches Institut, Aachen, Germany}}
\newcommand{\INSTJB}{\affiliation{Departamento de F\'isica At\'omica, Molecular y Nuclear, Universidad de Sevilla, 41080 Sevilla, Spain}}
\newcommand{\INSTFB}{\affiliation{University of Sheffield, Department of Physics and Astronomy, Sheffield, United Kingdom}}
\newcommand{\INSTDI}{\affiliation{University of Silesia, Institute of Physics, Katowice, Poland}}
\newcommand{\INSTBB}{\affiliation{Sorbonne Universit\'e, CNRS/IN2P3, Laboratoire de Physique Nucl\'eaire et de Hautes Energies (LPNHE), Paris, France}}
\newcommand{\INSTEH}{\affiliation{STFC, Rutherford Appleton Laboratory, Harwell Oxford,  and  Daresbury Laboratory, Warrington, United Kingdom}}
\newcommand{\INSTCH}{\affiliation{University of Tokyo, Department of Physics, Tokyo, Japan}}
\newcommand{\INSTBJ}{\affiliation{University of Tokyo, Institute for Cosmic Ray Research, Kamioka Observatory, Kamioka, Japan}}
\newcommand{\INSTCG}{\affiliation{University of Tokyo, Institute for Cosmic Ray Research, Research Center for Cosmic Neutrinos, Kashiwa, Japan}}
\newcommand{\INSTHF}{\affiliation{Tokyo Institute of Technology, Department of Physics, Tokyo, Japan}}
\newcommand{\INSTGI}{\affiliation{Tokyo Metropolitan University, Department of Physics, Tokyo, Japan}}
\newcommand{\INSTHG}{\affiliation{Tokyo University of Science, Faculty of Science and Technology, Department of Physics, Noda, Chiba, Japan}}
\newcommand{\INSTF}{\affiliation{University of Toronto, Department of Physics, Toronto, Ontario, Canada}}
\newcommand{\INSTB}{\affiliation{TRIUMF, Vancouver, British Columbia, Canada}}
\newcommand{\INSTDJ}{\affiliation{University of Warsaw, Faculty of Physics, Warsaw, Poland}}
\newcommand{\INSTDH}{\affiliation{Warsaw University of Technology, Institute of Radioelectronics and Multimedia Technology, Warsaw, Poland}}
\newcommand{\INSTIJ}{\affiliation{Tohoku University, Faculty of Science, Department of Physics, Miyagi, Japan}}
\newcommand{\INSTFD}{\affiliation{University of Warwick, Department of Physics, Coventry, United Kingdom}}
\newcommand{\INSTGH}{\affiliation{University of Winnipeg, Department of Physics, Winnipeg, Manitoba, Canada}}
\newcommand{\INSTEA}{\affiliation{Wroclaw University, Faculty of Physics and Astronomy, Wroclaw, Poland}}
\newcommand{\INSTHE}{\affiliation{Yokohama National University, Department of Physics, Yokohama, Japan}}
\newcommand{\INSTH}{\affiliation{York University, Department of Physics and Astronomy, Toronto, Ontario, Canada}}

\INSTHD
\INSTEE
\INSTFE
\INSTGA
\INSTI
\INSTGB
\INSTFG
\INSTFH
\INSTJA
\INSTEF
\INSTIE
\INSTEG
\INSTHJ
\INSTDG
\INSTCB
\INSTIB
\INSTED
\INSTJC
\INSTEC
\INSTHH
\INSTEI
\INSTGF
\INSTBE
\INSTBF
\INSTBD
\INSTEB
\INSTHI
\INSTJD
\INSTHA
\INSTID
\INSTIF
\INSTCC
\INSTCD
\INSTEJ
\INSTII
\INSTBA
\INSTFC
\INSTFI
\INSTIH
\INSTHB
\INSTCE
\INSTDF
\INSTFJ
\INSTGJ
\INSTCF
\INSTGG
\INSTIC
\INSTGC
\INSTFA
\INSTE
\INSTGD
\INSTHC
\INSTBC
\INSTJB
\INSTFB
\INSTDI
\INSTBB
\INSTEH
\INSTCH
\INSTBJ
\INSTCG
\INSTHF
\INSTGI
\INSTHG
\INSTF
\INSTB
\INSTDJ
\INSTDH
\INSTIJ
\INSTFD
\INSTGH
\INSTEA
\INSTHE
\INSTH

\author{K.\,Abe}\INSTBJ
\author{N.\,Akhlaq}\INSTFA
\author{R.\,Akutsu}\INSTCB
\author{A.\,Ali}\INSTGH\INSTB
\author{S.\,Alonso Monsalve}\INSTEF
\author{C.\,Alt}\INSTEF
\author{C.\,Andreopoulos}\INSTFC
\author{M.\,Antonova}\INSTEC
\author{S.\,Aoki}\INSTCC
\author{T.\,Arihara}\INSTGI
% ...

\collaboration{The T2K Collaboration}\noaffiliation

\date{\today}

\begin{abstract}
We report an updated measurement of the $\nu_{\mu}$-induced, and the first measurement of the $\bar{\nu}_{\mu}$-induced coherent charged pion production cross section on $^{12}\textup{C}$ nuclei in the T2K experiment. This is measured in a restricted region of the final-state phase space for which $p_{\mu,\pi} > 0.2$~GeV, $\cos(\theta_{\mu}) > 0.8$ and $\cos(\theta_{\pi}) > 0.6$, and at a mean (anti)neutrino energy of 0.85~GeV using the T2K near detector. The measured $\nu_{\mu}$ CC coherent pion production flux-averaged cross section on $^{12}\textup{C}$ is $(2.98 \pm 0.37 \textup{(stat.)} \pm 0.31\textup{(syst.)}\,\substack{ +0.49 \\ -0.00 }\,\mathrm{ (Q^2\,model)})  \times 10^{-40}~\mathrm{cm}^{2}$. The new measurement of the $\bar{\nu}_{\mu}$-induced cross section on $^{12}\textup{C}$ is $(3.05 \pm 0.71 \textup{(stat.)} \pm 0.39 \textup{(syst.)}\,\substack{ +0.74 \\ -0.00 }\,\mathrm{(Q^2\,model)}) \times 10^{-40}~\mathrm{cm}^{2}$. 
The results are compatible with both the NEUT 5.4.0 Berger-Sehgal (2009) and GENIE 2.8.0 Rein-Sehgal (2007) model predictions.
\end{abstract}

\maketitle

\section{Introduction}

Charged current coherent pion production in (anti)neutrino-nucleus
scattering, $\overset{\scriptscriptstyle(-)}{\nu}_{\mu} + A \rightarrow \mu^{-(+)} + \pi^{+(-)} + A$,
is a process in which a neutrino scatters coherently off a target nucleus. This process leaves the nucleus in its ground state, with the $W$-boson fluctuating to a charged meson (usually a pion) in the final state. No quantum numbers are exchanged with the nucleus and the magnitude of the square of the four-momentum transfer to the nucleus, denoted as $\left| t \right|$, must be small to maintain coherence. The interaction results in an unchanged nucleus, a lepton and pion in the final state 
and no other particles.

\begin{figure}[h]
\includegraphics[width=0.5\columnwidth]{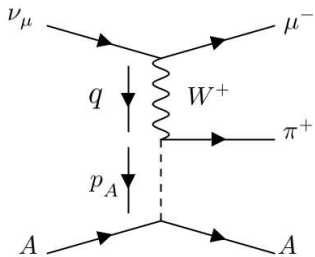}
\caption{\label{fig:coherentFD:PCAC} Feynman diagram for coherent charged pion production from a neutrino off a nucleus. This is specific to the PCAC class of models. The square of the magnitude of the 4-momentum transfer to the nucleus is $|p_{A}|^{2} = \lvert q - p_{\pi}\rvert^{2} = \lvert t \rvert$.}
\end{figure}

The most common theoretical description of this process is based on Adler's partially conserved axial vector current (PCAC) theorem\cite{Adler}, which connects the forward scattering amplitude (where the square of the 4-momentum transferred to the hadronic system, $-q^{2} = Q^{2}$ is equal to zero) with the divergence of the axial current. This in turn is estimated from the elastic pion-nucleus scattering cross section. The coherent neutrino (and antineutrino) scattering cross section at $Q^{2} = 0$ can then be written as
\begin{equation}
\frac{d^3\sigma_{coh}}{dQ^{2}\,dy\, d|t|}\biggr \vert_{Q^{2}=0} = \frac{G_{F}^{2}}{2 \pi^{2}}  f_{\pi}^{2} \frac{1-y}{y} \frac{d\sigma( \pi A \rightarrow \pi A)}{d|t|},
\label{eq:Adler}
\end{equation}
where $y = E_{\pi} / E_{\nu}$ with $E_{\pi}$ and $E_{\nu}$ being the energy of the pion and neutrino, respectively and
$f_{\pi}$ is the pion decay constant.
A  Feynman diagram for this process is shown in Fig.~\ref{fig:coherentFD:PCAC}. This cross section is then extrapolated to higher $Q^{2}$. PCAC models use a variety of methods for the $Q^{2}$ extrapolation, as well as different approaches to characterise pion-nucleus scattering. The most common model currently used by Monte Carlo (MC) neutrino event generators~\cite{Andreopoulos:2009rq, NEUT, NUWRO, Buss:2011mx} has been the Rein-Sehgal (RS) model~\cite{ReinSehgalCoh}. This uses pion-proton and pion-deuterium data along with a simple $A$-scaling and ad hoc description for nuclear effects like pion absorption. It was developed for neutrino energies above approximately 5\,GeV where the mass of the final state lepton has minimal effect. 
The newer Berger-Sehgal (BS) model~\cite{BergerSehgalCoh} updates this approach with the use
of pion-carbon scattering data, which features a significant reduction in the resonance peak. The two models are identical for pion kinetic energies above 1.5 GeV, and employ a similar A-scaling technique. Different characterizations of the pion scattering data (pion-proton for RS, and pion-carbon for BS) in various generators can account for observed differences in their model predictions.
Independent MC simulation sets using the NEUT 5.4.0~\cite{NEUT} Berger-Sehgal (2009) and GENIE 2.8.0~\cite{Andreopoulos:2009rq} Rein-Sehgal (2007) model implementations were used for this analysis. 

The most recent charged current coherent production cross section measurements at high neutrino energies (above 7\,GeV) were made in the 1980-1990s~\cite{E632:1992ydt,Ammosov,marage1989coherent,Vilain:1993sf,Grabosch:1985mt,Willocq:1992fv} and were found to agree with the Rein-Sehgal model. The discovery of neutrino oscillations~\cite{Super-Kamiokande:1998kpq,SNO:2002tuh,K2K:2002icj,KamLAND:2004mhv} refocused the neutrino community on lower energies where a scarcity of data on this interaction mode existed. At neutrino energies around 0.5--2.0\,GeV, upper limits of the cross section from K2K~\cite{Hasegawa:2005td} and SciBooNE~\cite{Hiraide:2008eu} and a measurement by T2K~\cite{T2K:2016soz}
were significantly lower than that of the Rein-Sehgal model, but agreed with Berger-Sehgal.
The MINERvA experiment, which operated at neutrino energies of 1.5--20\,GeV, was the first to report measurements of differential cross sections in the variables $Q^{2}$, $E_{\pi}$ and $\theta_{\pi}$~\cite{MINERvA:2017ipy} on a set of different nuclear targets~\cite{MINERvA:2022esg}. The collaboration found that the measured total cross sections agreed with the predictions from both models, but that 
the observed differential cross sections in the pion angle and energy variables showed
an excess in the forward region with respect to model predictions. 
The MINERvA experiment also made the first observation of coherent kaon production~\cite{MINERvA:2016cun} and neutral current coherent neutral pion production in an antineutrino beam~\cite{MINERvA:2015slz}. Neutral current coherent production of neutral pions was also measured by the MINOS~\cite{MINOS:2016yyz} and the NOvA~\cite{NOvA:2019bdw} collaborations.

This letter presents the first measurement of the antineutrino induced coherent pion production cross section on $^{12}\textup{C}$ at a mean neutrino energy of 0.85\,GeV. In addition, the previous T2K
measurement of neutrino-induced coherent pion production~\cite{T2K:2016soz} is updated by doubling the size of the available data set and updating the systematic uncertainty estimates.

\section{The T2K Experiment}

The Tokai-to-Kamioka (T2K) experiment is the second generation experiment in the long-baseline neutrino oscillation program operating in Japan. T2K established the oscillation from muon neutrinos to electron neutrinos~\cite{T2K:2013ppw} and is investigating  charge-parity violation in the leptonic sector as well as measuring precisely other oscillation parameters~\cite{T2K:2021xwb}.
Details of the T2K experiment can be found in Ref.~\cite{T2K:2011qtm}.
Although the focus of T2K is on neutrino flavour oscillation studies, T2K has also  studied (anti)neutrino-nucleus interactions in the few hundreds of MeV to few GeV neutrino energy 
range extensively (for example, see References~\cite{ 
T2K:2016soz, 
T2K:2016cbz, 
T2K:2018lnf, 
T2K:2019ddy, 
T2K:2020txr, 
T2K:2020jav, 
T2K:2020lrr, 
T2K:2021naz}).

\subsection{The muon (anti)neutrino beam}

T2K employs the J-PARC neutrino beamline, as detailed in Ref.~\cite{T2K:2011qtm}, to generate an intense and near-pure beam of muon (anti)neutrinos. The (anti)neutrino beam is produced from the decay-in-flight of pions and kaons produced when a 30~GeV proton beam from the J-PARC Main Ring synchrotron is guided onto a cylindrical target consisting of disks of graphite evenly spaced along a length of 91.4~cm and a diameter of 2.6~cm. Current transformers, secondary emission monitors, and optical transition radiation monitors are used to monitor the intensity and profile of the proton beam before hitting the target. Further information about the beam is provided by a muon monitor which measures the intensity and direction of muons produced from the meson decays. A set of 14 scintillator-iron sampling calorimeter modules, each comprising 7.1 tonne of iron, and collectively referred to as the INGRID~\cite{Abe:2011xv} is used to measure and monitor the stability of neutrino intensity with better than 1\% precision and the neutrino beam direction with a precision better than 1~mrad.  

T2K employs a system of three magnetic horns to focus secondary charged particles, the parents of neutrinos, and defocus oppositely charged particles. The horn polarity determines whether the beam is configured in the muon (anti)neutrino, or (Reverse) Forward Horn Current mode, abbreviated as
(RHC) FHC, where the focusing horns in the beamline are operated with a current of (-250 kA) 250 kA.
The total exposure used in this study is  $11.54\times 10^{20}$ protons-on-target (POT) taken in FHC configuration from January 2010 to April 2017,
and $8.15\times 10^{20}$ POT taken in RHC configuration from June 2014 to May 2018.
The statistics in the FHC configuration have been doubled since the previous T2K publication~\cite{T2K:2016soz}, while the data in the RHC configuration is new.

T2K's approach to understand the neutrino beam is described in Ref.~\cite{T2K:2012bge}. T2K employs data-based tuning to precisely predict the flux. These data include the measured proton beam parameters from the beam monitors, hadron production data from the NA61/SHINE fixed target experiment at CERN's Super Proton Synchrotron, and the INGRID beam direction measurements. Compared to the previous publication on the coherent neutrino-nuclei interaction~\cite{T2K:2016soz}, this study implements an updated flux prediction~\cite{t2kCollaboration2021:5734307} to include measurements of $\pi^{\pm}$ production from the NA61/SHINE~\cite{NA61:2014lfx} experiment operating with a replica of the T2K target~\cite{NA61SHINE:2016nlf}. With this significant update, the flux uncertainty is reduced to approximately 5\% near the peak (around 600~MeV) of the neutrino spectrum, comparing to 8.5\% flux uncertainty in the previous measurement of coherent pion production. Along with the main flavour components ($\nu_{\mu}$ in FHC and $\overline{\nu}_{\mu}$ in RHC), the beam contains a small fraction of wrong-sign component ($\approx 5\%$ $\overline{\nu}_{\mu}$ in FHC and  $\approx 7\%$ $\nu_{\mu}$ in RHC) and intrinsic electron (anti)neutrino components ($\nu_e$ and $\overline{\nu}_e$) at a level less than 1\%~\cite{T2K:2012bge,t2kCollaboration2021:5734307,T2K:2014djs}.

The nominal Monte Carlo model used for this measurement is the NEUT 5.4.0~\cite{NEUT} neutrino event generator. Charged current coherent single pion production events are generated using the Berger-Sehgal\cite{BergerSehgalCoh} model. The backgrounds to this process are dominated by charged current resonant pion production (CC-RES)  and deep inelastic scattering (DIS). CC-RES processes are modelled using the Rein-Sehgal formalism\cite{Rein:1980wg} updated
to implement the effect of the final state charged lepton mass\cite{Berger:2007rq}, and using updated nucleon axial
form factors\cite{Graczyk:2014dpa}. Contributions from 17 baryon resonances are considered, with
the $\Delta(1232)$ being dominant, and interference terms between the resonances are taken into account. DIS interactions are modelled using the GRV98 parton distribution functions\cite{Gluck:1998xa} with low Q$^{2}$ corrections from Bodek and Yang\cite{Bodek:2003wc}.
CC-RES events are produced in the invariant hadronic mass region of $W<2$~GeV, with the DIS event production beginning in the invariant hadronic mass region of $W > 1.3$~GeV. In the overlap region, $ 1.3~\mathrm{GeV} < W < 2.0~\mathrm{GeV}$, a custom hadronisation model\cite{Hayato:2021heg} is used to interpolate between the two processes. Above an invariant hadronic mass of 2~GeV, hadronisation is managed by PYTHIA 5.7 and JETSET 7.4\cite{Sjostrand:1994kzr}. Final state interactions of hadrons as they propagate through the nuclear medium are modelled using a custom intranuclear cascade model\cite{Hayato:2021heg}.

\subsection{The T2K off-axis near detector complex}

The T2K near detector, referred to as ND280, is placed $280$~m from the proton interaction target at the same off-axis angle as the T2K far detector. The detector is intended to characterise the neutrino beam prior to oscillation and precisely measure the interactions of muon (anti)neutrinos and electron (anti)neutrinos with carbon and water. ND280, which is described in detail in Ref.~\cite{T2K:2011qtm}, consists of multiple sub-detector systems positioned inside a magnet producing a magnetic field of 0.2~T.
These systems include an upstream $\pi^0$ detector and a tracking detector containing three Time Projection Chambers (TPCs)~\cite{T2KND280TPC:2010nnd} interleaved with two Fine Grained tracking scintillator Detectors~\cite{T2KND280FGD:2012umz} (referred to as FGD1 and FGD2) constructed from plastic scintillator bars.
The FGDs provide the target mass for the neutrino interactions as well as fine grained tracking of charged particles from the interaction vertex.
The TPCs identify the type of charged particle and measure the momenta of particles leaving the FGDs. Each FGD weighs 1.1 tonnes and measures $1.84~\mbox{m (width)} \times 1.84~\mbox{m (height)} \times 0.37~\mbox{m (depth)}$. The upstream fine grained detector (FGD1) is constructed with fifteen interleaved plastic scintillator modules, each of which has two layers, segmented with 192 extruded plastic scintillator bars with $0.96~\mbox{cm}^{2}$ cross sectional area, oriented in the X and Y directions transverse to the neutrino beam direction.
The downstream fine grained detector, FGD2, contains six $2.5~\mbox{cm}$ thick layers of water to provide a water-enriched neutrino target, each surrounded by two XY scintillator modules. Only events with the neutrino vertex reconstructed in FGD1 were used in this study.

\section{Analysis Strategy}

\begin{figure*}[!htbp]
\includegraphics[width=0.4\textwidth]{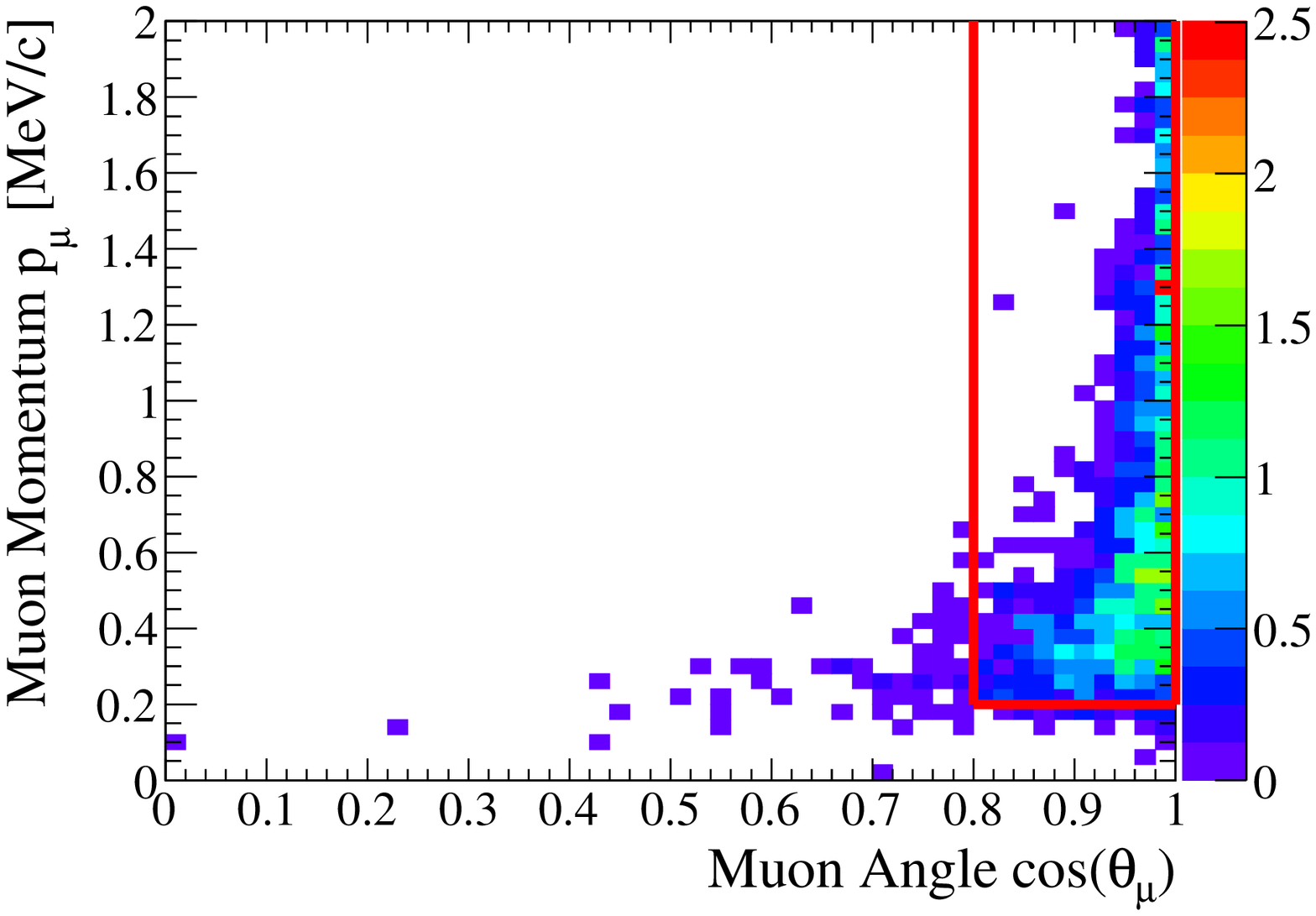}
 \includegraphics[width=0.4\textwidth]{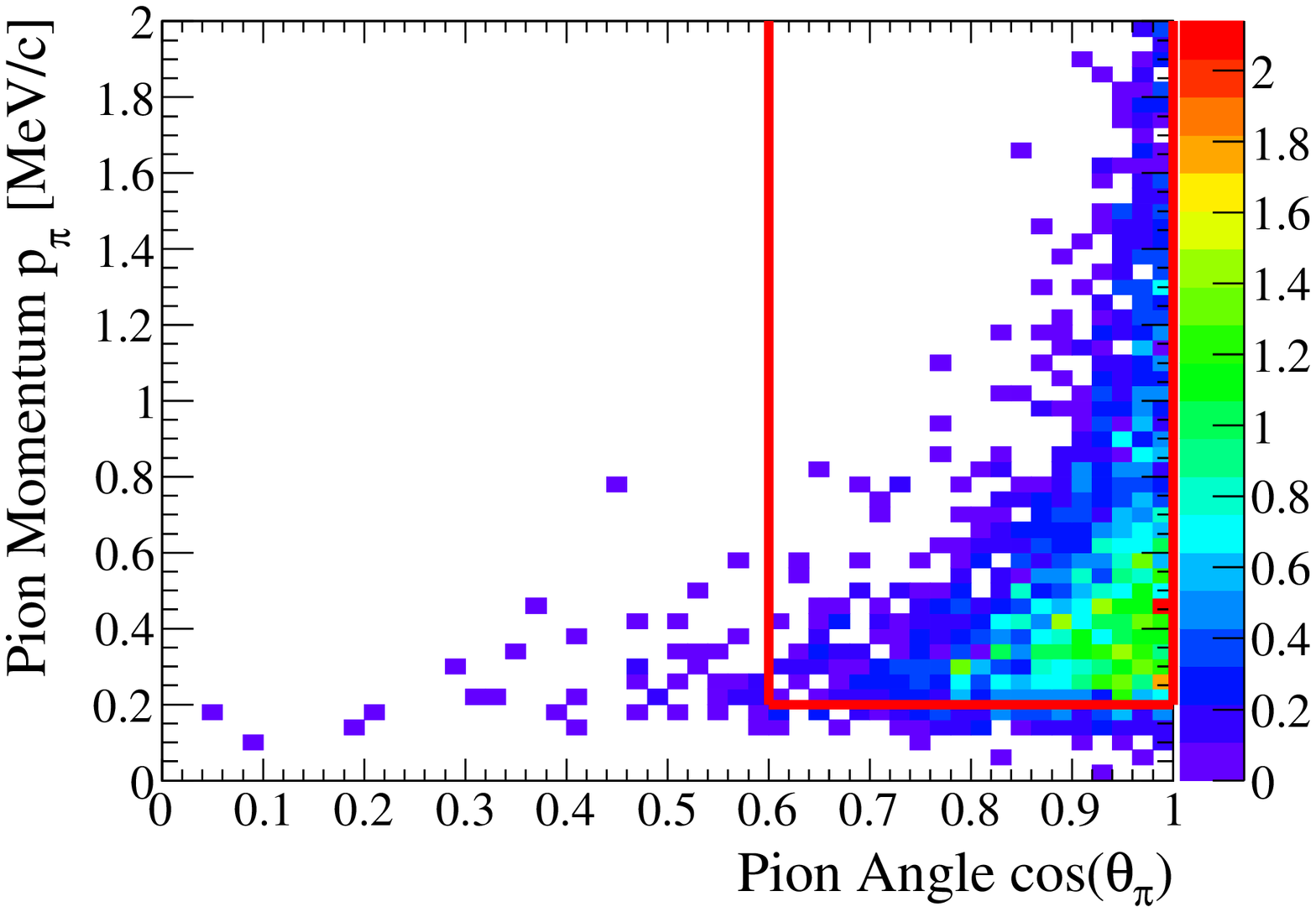}
 \includegraphics[width=0.4\textwidth]{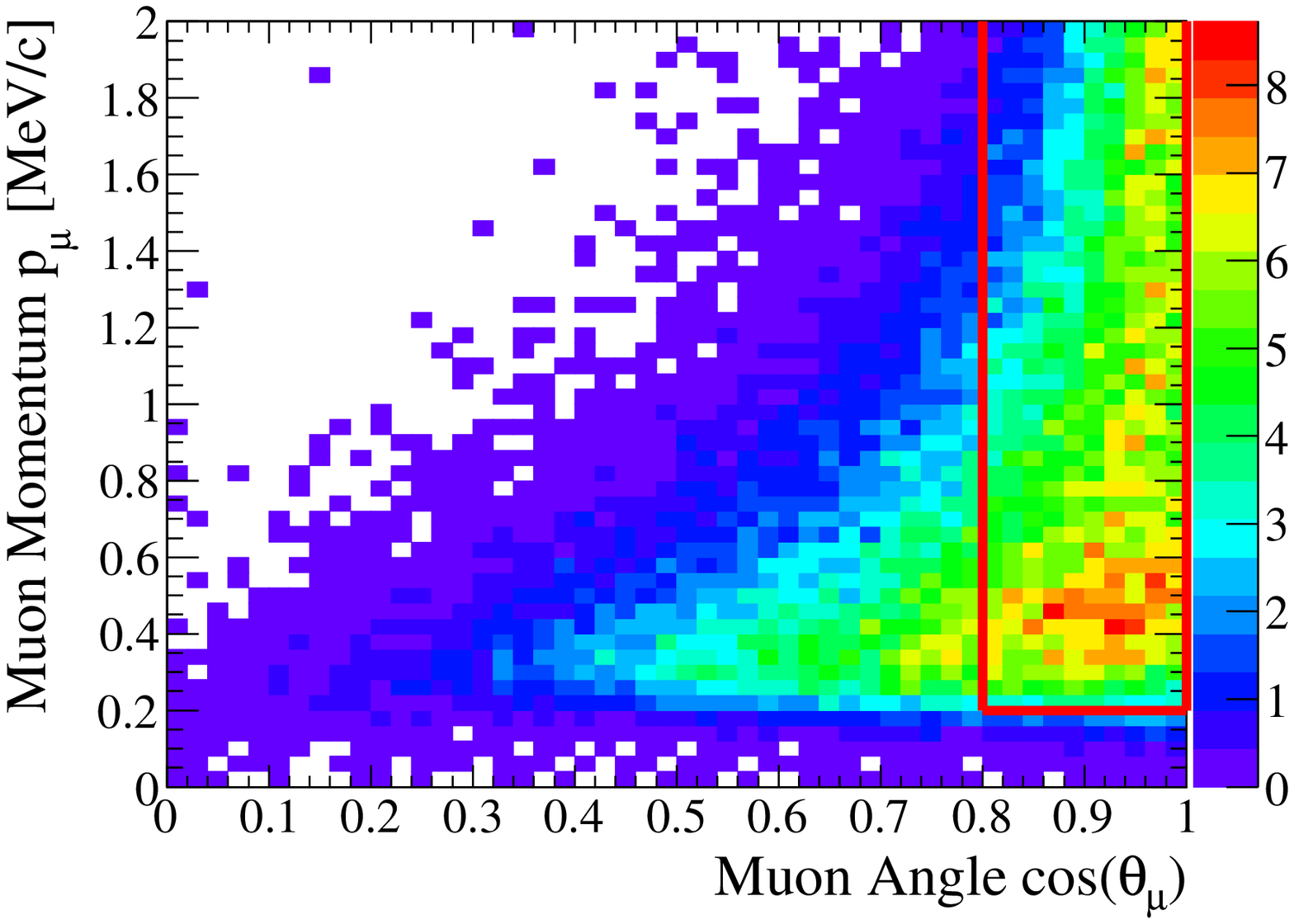}
 \includegraphics[width=0.4\textwidth]{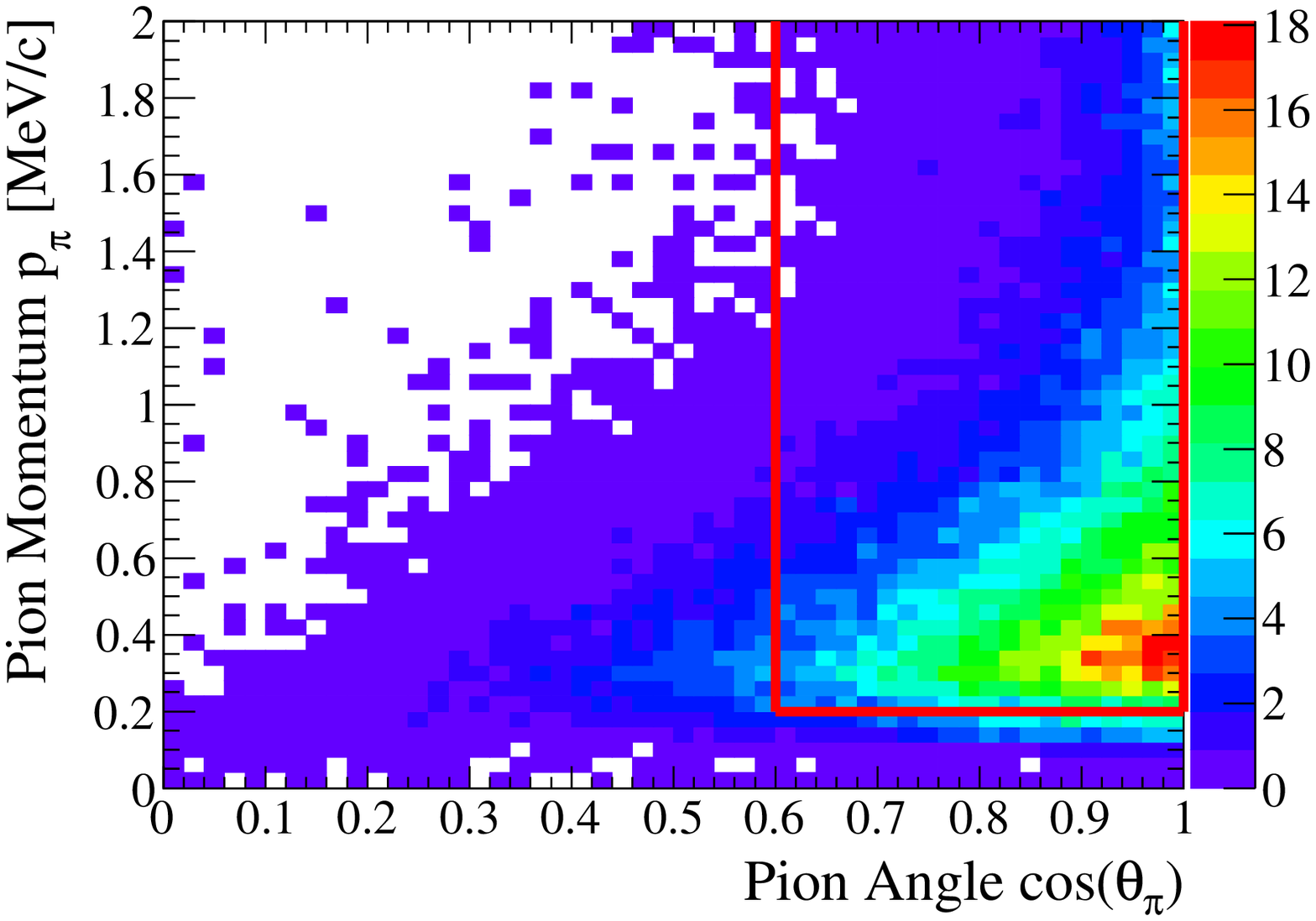}
  \caption{The muon (left) and pion (right) kinematic distributions for the $\nu _{\mu}$ coherent signal events (top) and for all signal and background events (bottom). The straight lines indicate the restricted phase space; note no upper bound is set for the muon and pion momentum. Based on the NEUT Monte Carlo, 82.9\% of the true COH events remain in the restricted phase space.}
  \label{fig:numu_ps_cut}
\end{figure*}

\begin{figure*}[!htbp]
\includegraphics[width=0.4\textwidth]{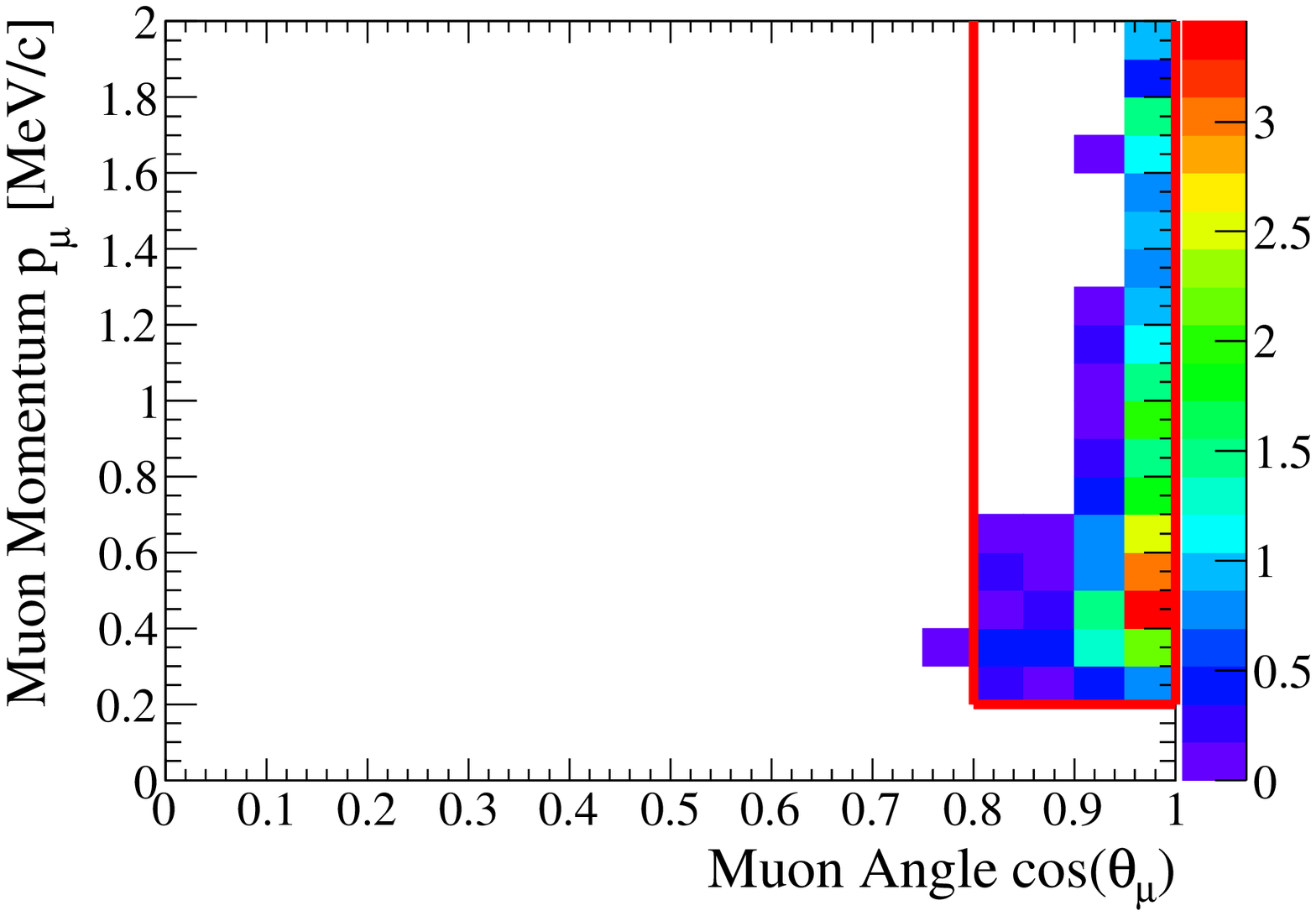}
 \includegraphics[width=0.4\textwidth]{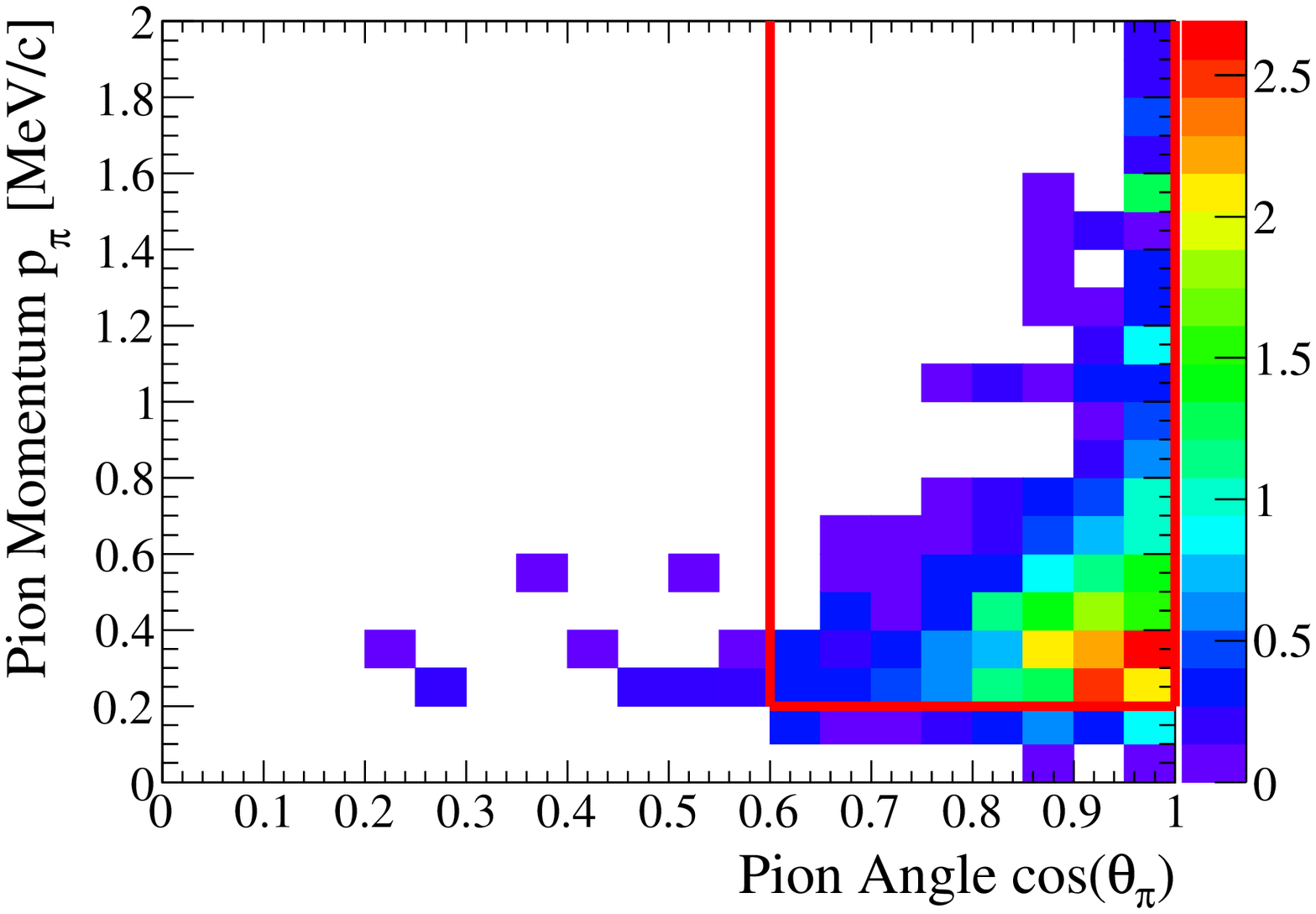}
 \includegraphics[width=0.4\textwidth]{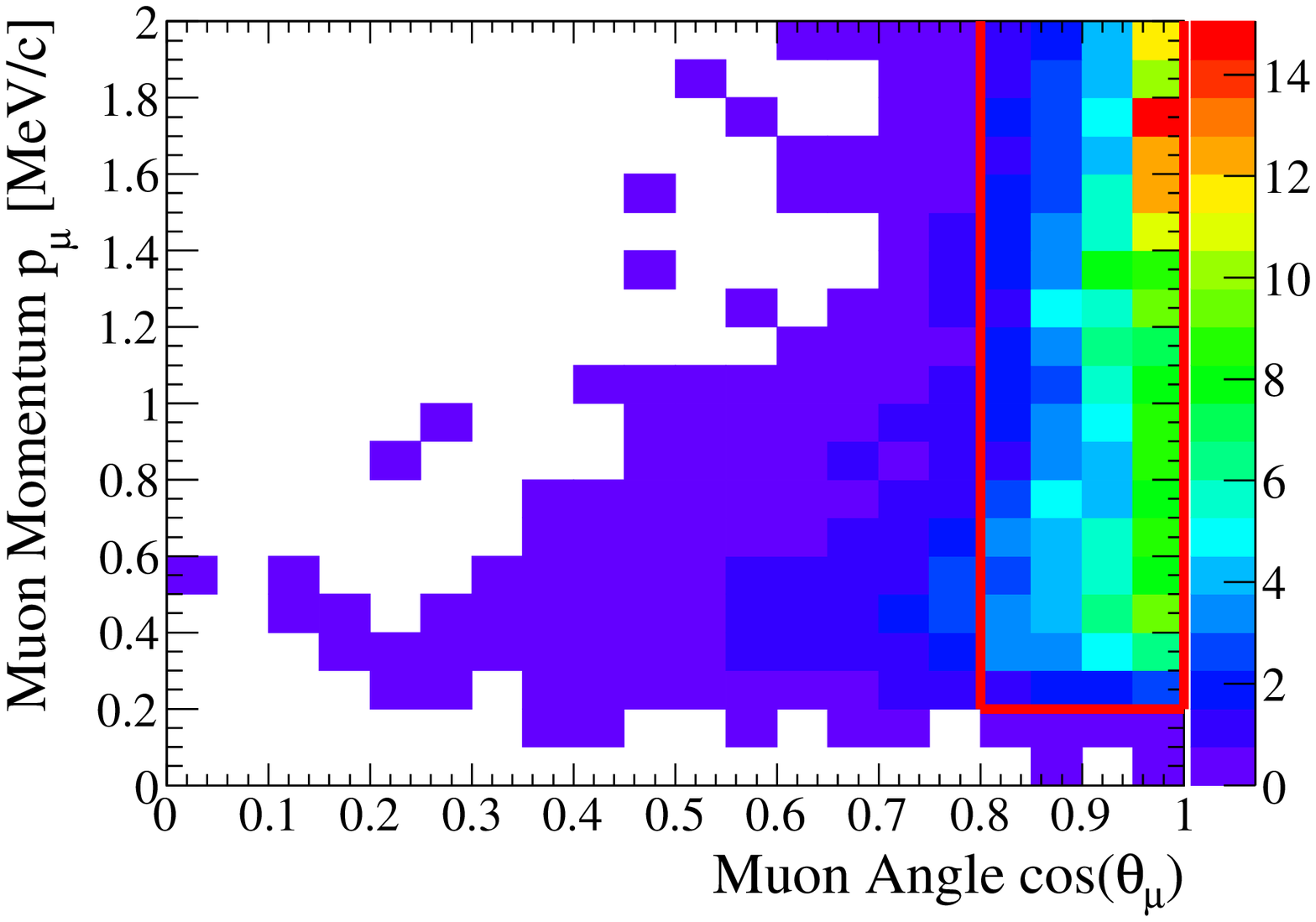}
 \includegraphics[width=0.4\textwidth]{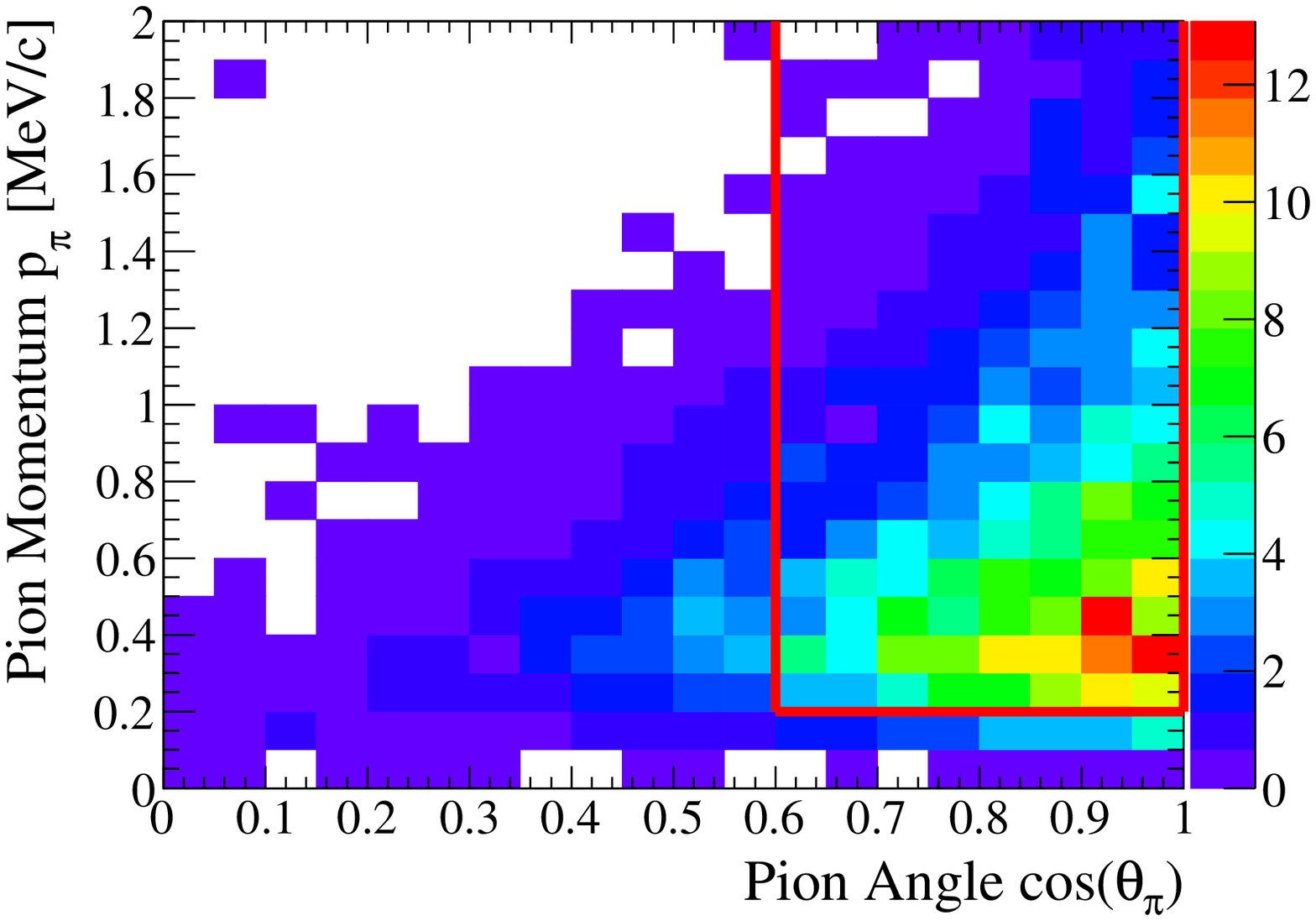}
  \caption{The muon (left) and pion (right) kinematic distributions for the $\bar{\nu} _{\mu}$ coherent signal events (top) and for all signal and background events (bottom). The straight lines indicate the restricted phase space; note no upper bound is set for the muon and pion momentum. Based on the NEUT Monte Carlo, 83.5\% of the true COH events remain in the restricted phase space.}
  \label{fig:numub_ps_cut}
\end{figure*}

The signature of the coherent charged pion production process is one muon and one charged pion, both forward-going in the detector.
The detector's charged particle tracking and identification efficiency restricts the measurable phase space of the muon and pions to kinematic ranges:
\begin{itemize}
    \item $p_{\mu}$ $>$ 0.2 GeV and $\textup{cos}(\theta_{\mu})$ $>$ 0.8,
    \item $p_{\pi}$ $>$ 0.2 GeV and $\textup{cos}(\theta_{\pi})$ $>$ 0.6.
\end{itemize}
The muon and pion momenta and angular distributions are shown in Fig.~\ref{fig:numu_ps_cut} and~\ref{fig:numub_ps_cut}, along with an indication of the size of the restricted phase space. 
The selection of these constraints is based on the performance in the NEUT Monte Carlo and includes (82.9\%)83.5\% of all coherent events in the (RHC)FHC beam mode.

These criteria are based on the charged current single pion production selection as described in~\cite{T2K:2021xwb}. Beyond that, the target nucleus is left intact, so no additional hadronic activity should be detected in the region around the interaction vertex.
Low vertex activity (VA), defined as the energy deposited in a $5 \times 5 \times 5$ (approximately $(5~\mbox{cm})^{3}$) volume of scintillator around the vertex position, is required. A further restriction is applied to the $|t|$ distribution, which can be calculated from the muon and pion kinematic variables:

\begin{widetext}
\begin{equation}
  \left | t \right |  =  \biggl(\sum_{i=\mu,\pi}\left(E_{i}-\left | \overrightarrow{P_{i}} \right |\cos(\theta_{i})\right)\biggr)^2 +  \left | \sum _{i=\mu,\pi}  \left | \overrightarrow{P_{i}} \right | \bigl( \widehat{e}_{i}- \cos(\theta_{i})\widehat{e}_{\nu}\bigr)\right |^{2} \text{,}
  \label{eq:strategy_t}
\end{equation}
\end{widetext}
where $E_{i}$ is the energy, $\overrightarrow{P_{i}}$ is the momentum, $\widehat{e}_{i}$ is the direction unit vector, and $\theta_{i}$ is the scattering angle of the muon and pion in the event, $\widehat{e}_{\nu}$ is the neutrino direction unit vector.

\begin{figure*}[!htbp]
 \includegraphics[width=0.4\textwidth]{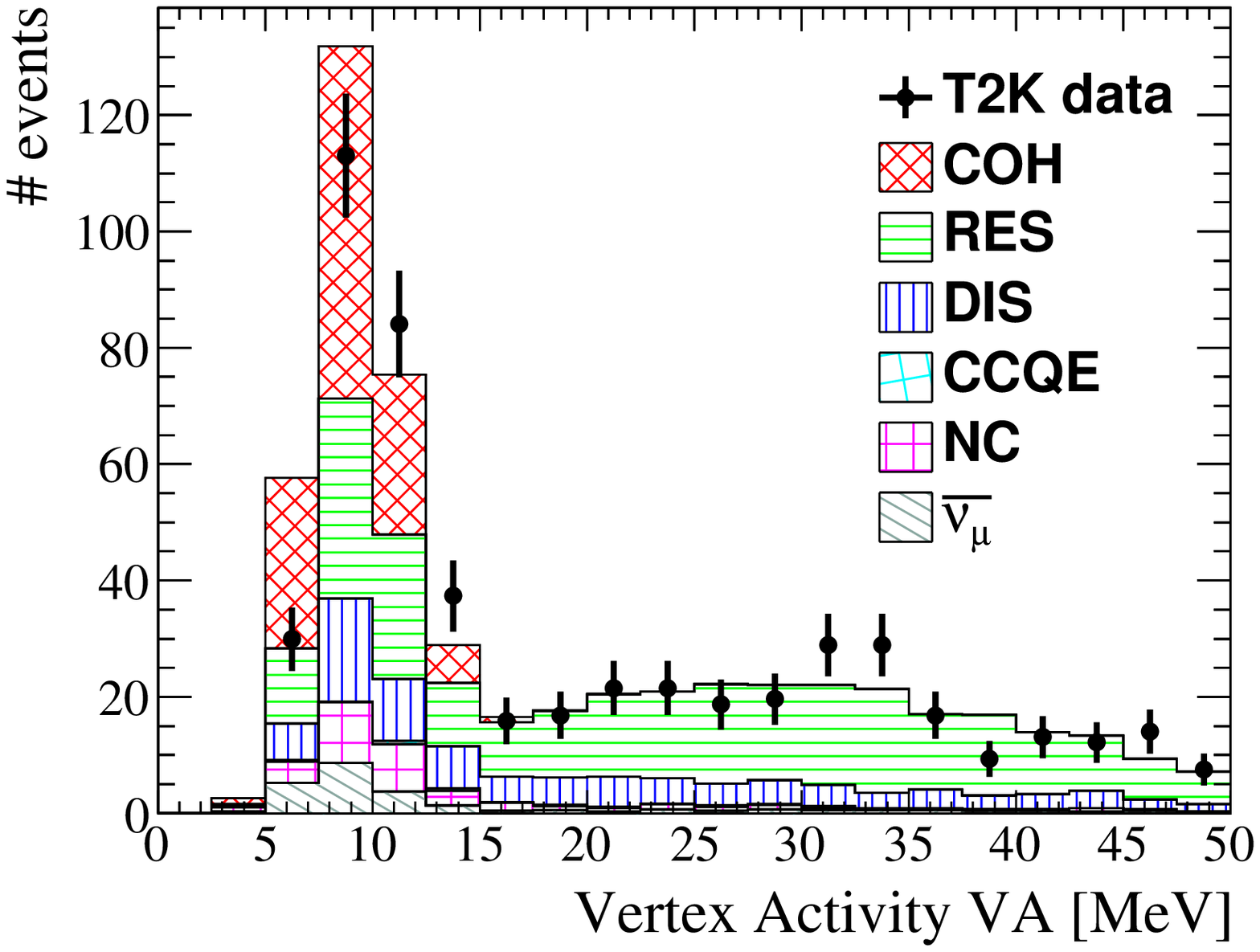}
 \includegraphics[width=0.4\textwidth]{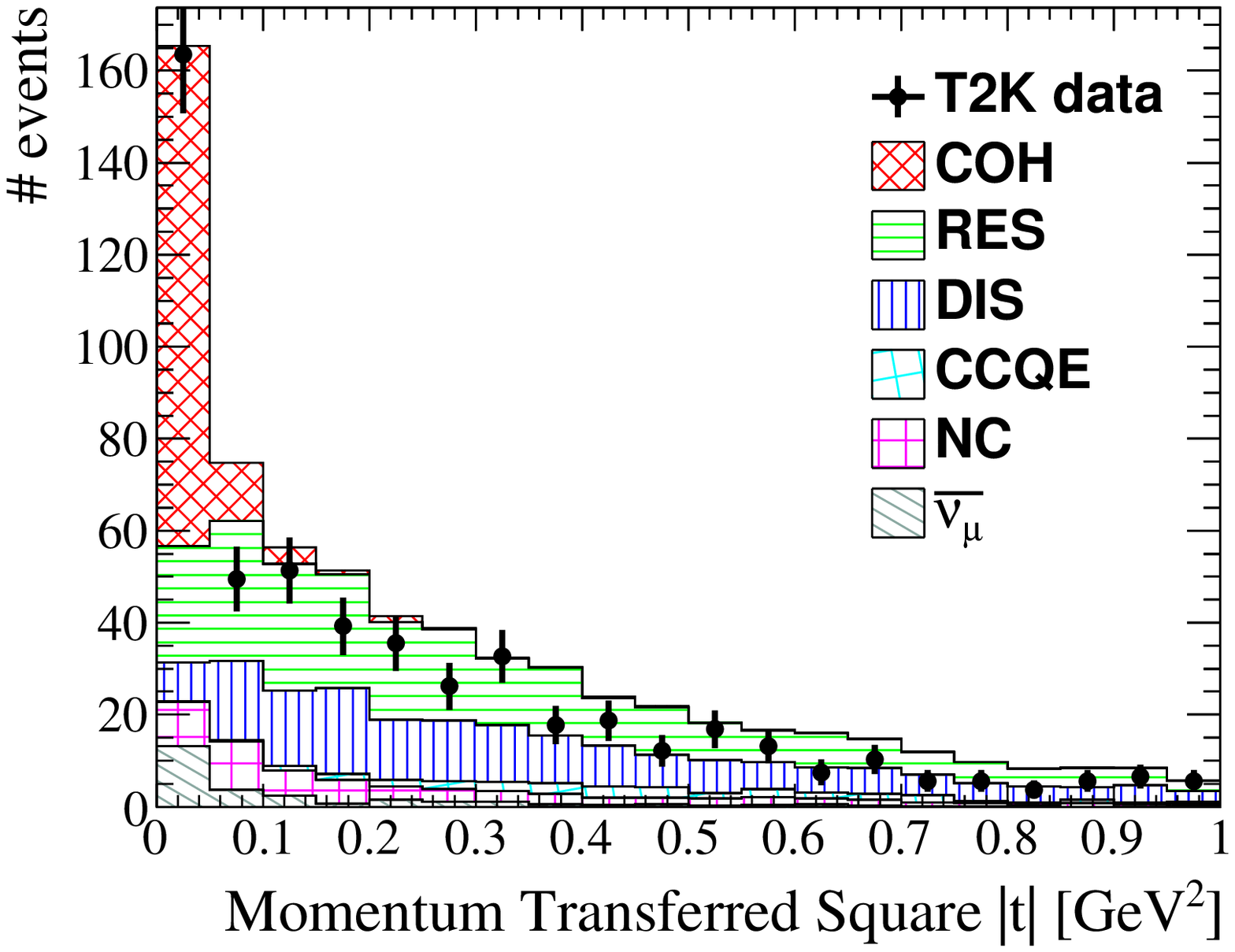}
 \includegraphics[width=0.4\textwidth]{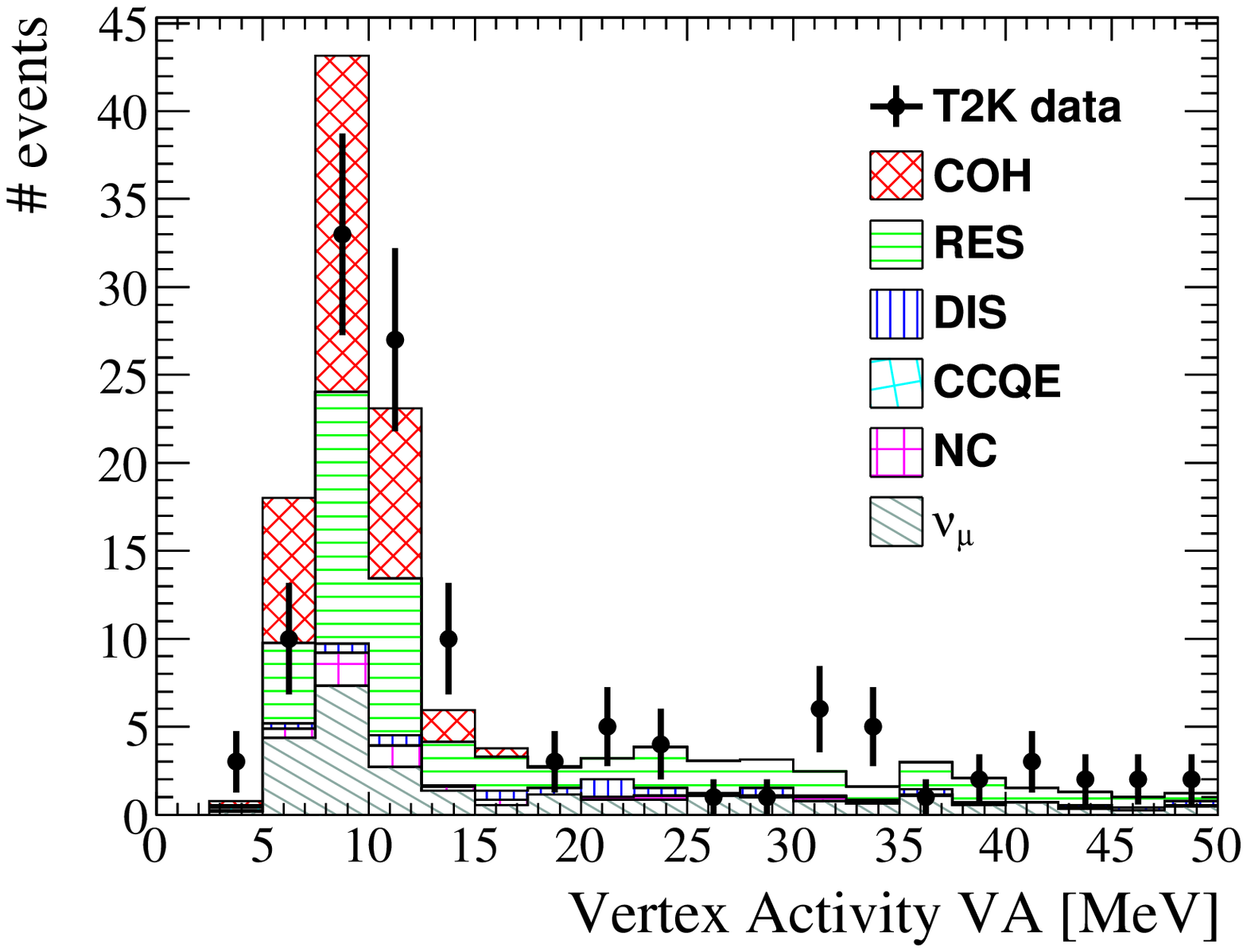}
 \includegraphics[width=0.4\textwidth]{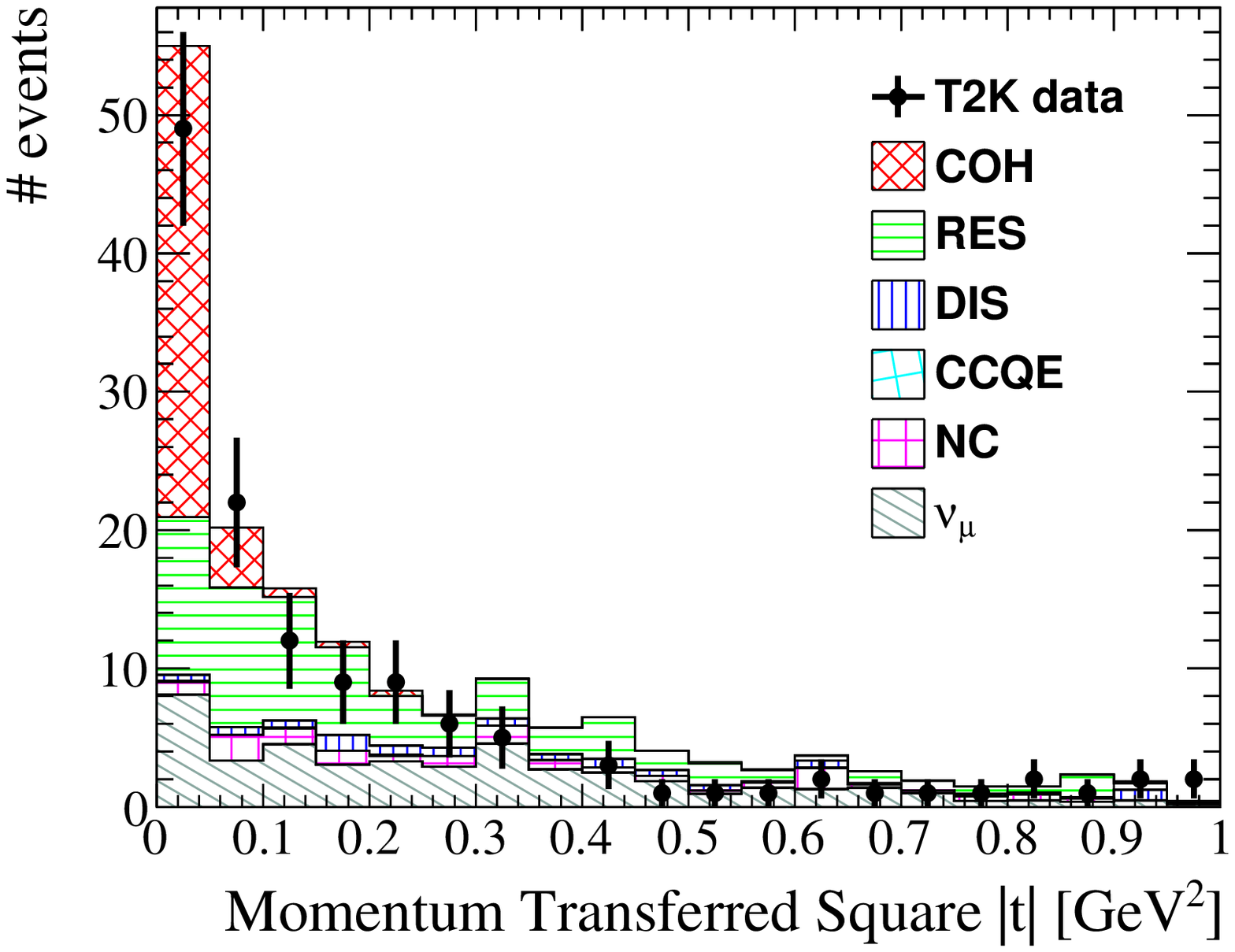}
  \caption{The $\nu _{\mu}$ (top) and $\bar{\nu}_{\mu}$ (bottom) COH selection, containing one muon and one charged pion, are shown in the VA (left) and $|t|$ (right) variables. The signal COH events for both selections concentrate in the low VA (less than $\SI{15}{MeV}$) and low $|t|$ (less than $\SI{0.15}{\square\GeV}$) region. The events with VA greater than $\SI{15}{MeV}$ are rejected. The events with $|t|$ greater than $\SI{0.15}{\square\GeV}$ are used for background control.
  The stacked histogram shows the pre-fit simulation overlaid by data.}
  \label{fig:selection_va_t}
\end{figure*}

The VA and $|t|$  distributions in data and in nominal simulation and for both $\nu _{\mu}$ and $\bar{\nu}_{\mu}$ after selecting events containing exactly one muon and one charged pion  
are shown in  Fig.~\ref{fig:selection_va_t}. The simulation shows that the coherent signal events (labelled COH here and below) are all concentrated in the low VA and low $|t|$ region as expected.
Background events with VA greater than $\SI{15}{MeV}$ or $|t|$ greater than $0.15~(\mbox{GeV})^{2}$ are removed from the event sample. 
The MC prediction for the signal purity and selection efficiency in the $\nu _{\mu}$ sample derived from the nominal MC are 41.6\% and 47.3\% respectively.
The major sources of background events are resonant pion production (28.0\%) and deep inelastic scattering (14.2\%). The signal purity in the $\bar{\nu}_{\mu}$ sample is 42.2\% with a 30.8\% selection efficiency. The major sources of background events in the $\bar{\nu}_{\mu}$ sample are resonant pion production (33.0\%) and interactions of $\nu _{\mu}$ contamination in the $\bar{\nu}_{\mu}$ beam (18.2\%). These values are all based on the nominal NEUT MC predictions.

The cross section was extracted using a binned likelihood fitter which is described in detail in Ref.~\cite{T2K:2021naz,T2K:2023zaf}.
The inputs to the fitter include templates that map the signal in each bin of true kinematic space to the associated bins in reconstructed kinematic space. Template weights assigned to each true bin are varied by the fitter, and the weighted templates are summed in bins in reconstructed space to generate the signal prediction at each fit point. In addition to these
signal normalisation parameters, there is a set of nuisance parameters associated with uncertainties in the cross section, flux and detector models, which change the shape of the templates as well as the shape and normalisation of the backgrounds.
The signal normalisation parameters are allowed to float freely in the fit, whilst the nuisance parameters are constrained by external measurements which are introduced to the procedure via a pre-fit covariance matrix.
The fitter uses the MINUIT2~\cite{James:1994vla} minimisation routine, MIGRAD, to maximise an extended likelihood. The post-fit parameter values are then used to calculate the number of signal events in the true kinematic space which best fits the data. The post-fit covariance between the fit parameters is estimated by calculating the error matrix at the best fit point using the HESSE algorithm~\cite{James:1994vla}.

In this measurement there is one signal template with (two)three bins in reconstructed parameter space corresponding to the (RHC)FHC mode. In both modes, the template contains a single CC-COH bin recording the number of selected signal events in the constrained reconstructed phase space.
The template also contains background-dominated sideband bins which are used to constrain the nuisance parameters describing the dominant RES and DIS backgrounds. In both beam modes, a sideband contains the sample of events having a reconstructed $|t|$ outside the CC-COH range. In the case of the FHC mode measurement, a three-track sample is used to constrain the DIS events. This was unnecessary in the RHC mode measurement since the fractional contribution of DIS events in the MC predicted background was less than 2\%.

The neutrino and antineutrino coherent pion production cross sections  are independently extracted by calculating 
\begin{equation}
\sigma_{\textup{FGD1}} =\frac{N_{\textup{FGD1}}}{\epsilon \cdot T_{\textup{FGD1}} \cdot \Phi }.
\label{eq:xsec_fgd1}
\end{equation}
where $N_{\textup{FGD1}}$  is the number of COH events obtained by the likelihood fitter, $\epsilon$ is the detector efficiency to select the COH events, $T_{\textup{FGD1}}$ is the number of target nuclei in FGD1, and $\Phi$ is the integrated muon (anti)neutrino flux. 
Each of these variables are functions of the fit parameters. These parameters are randomly sampled from the post-fit covariance of all fit parameters, and the cross sections calculated.
The resulting distribution of the cross section yields the final cross section uncertainty.

The value of $\sigma_\text{FGD}$ calculated at the best-fit point represents the average coherent cross section per non-hydrogen atom in the FGD1 fiducial volume (FV).
The hydrogen atoms are not considered here, because there is no coherent interaction on single protons.
For the purpose of this analysis, we consider the diffractive pion production on hydrogen~\cite{Rein:1986cd} as a background process.
A simulation~\cite{Andreopoulos:2009rq} showed that contamination of these diffractive events in the selected data is negligible. This conclusion is supported by the previous T2K coherent analysis~\cite{T2K:2016soz} which employed an independent method, developed by the MINERvA collaboration\cite{MINERvA:2017ipy}, to estimate the size of the diffractive component. This method showed that the fractional contribution of diffractive events in the coherent event sample was at most 5\% or, on average, (4)8 events in the current (RHC)FHC coherent event sample.
The relative elemental composition in the FV is shown in Table~\ref{tab:fgd1_fraction}.
The flux-shape uncertainties are handled in accordance with the prescription for ``the second approach'' discussed in detail in~\cite{Koch2020}.
The cross section is reported for the nominal T2K off-axis flux prediction~\cite{t2kCollaboration2021:5734307}.
All effects of flux-shape variations are covered by the uncertainties provided in this result.

\begin{table*}[!htbp]
\centering
\caption{Number of Target fractional composition of the FGD1 detector excluding hydrogen.}
\label{tab:fgd1_fraction}
\begin{tabular}{rccccc} 
\hline 
Element
& C        & O        & Ti       & Si       & N  \\ \hline
Atomic Mass Number ($A$)
& 12       & 16       & 48       & 28       & 14 \\
Fractional Composition ($f$) (\%)
& 95.83    & 3.09     & 0.46     & 0.48     & 0.14 \\
Relative uncertainty of Fractional Composition (\%)
& 0.5      & 1.3      & 16.6     & 19.7     & 39 \\ \hline
\end{tabular}
\end{table*}

The majority of atoms in FGD1 are carbon.
For easier comparison with other experiments the COH cross section on a carbon nucleus is calculated.
This can be achieved by using a scaling function, $F(A)$, which can account for elements of atomic mass number $A$:
\begin{equation}
    \sigma_\text{FGD1} =\sigma_\text{C} \sum_{i} f_{i} \frac{F(A_{i})}{F(A_{C})}, \quad i = \text{C, O, Ti, Si, N.}
\end{equation}
where $f_{i}$ represents the fractional composition of a given element. In this study the scaling function proposed in~\cite{ReinSehgalCoh}, $F(A) = A^{1/3}$, is used. Results using an alternative scaling function, $F(A) = A^{2/3}$, were also calculated. The difference in the results is small compared to the measurement uncertainties shown in Table~\ref{tab:results}.

A number of tests with simulated data were carried out to validate the neutrino cross section extraction methodology and to identify potential biases caused by neutrino interaction mismodeling. 
These included different increases and suppression of deep inelastic scattering, coherent, and resonance interaction modes, shifts in the high energy part of the neutrino flux, and the use of a completely different event generator, GENIE.
All simulated data studies showed that the analysis performed as expected.

\section{Results}

\begin{figure*}[!htbp]
\centering
\includegraphics[width=0.9\textwidth]{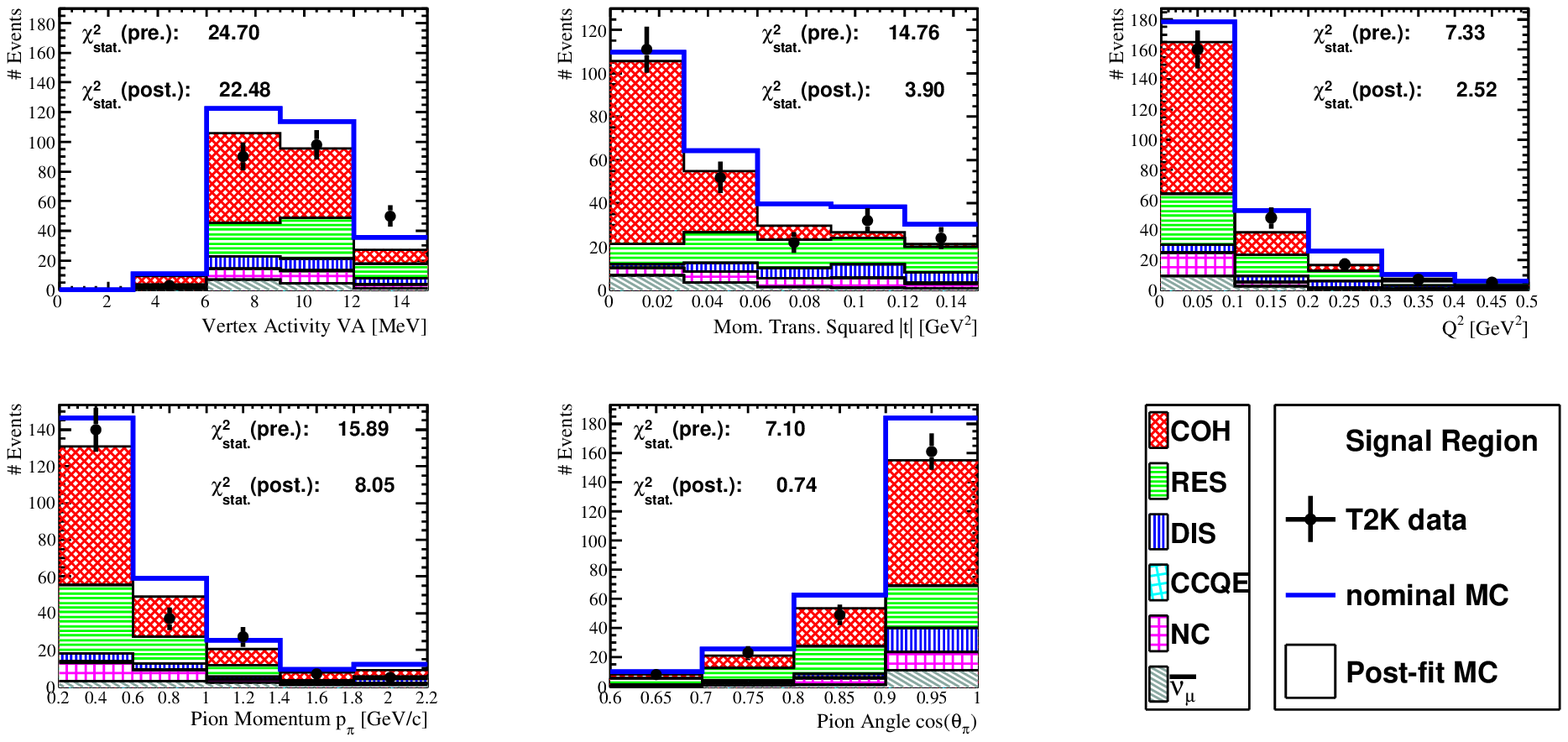}
\caption{$\nu_{\mu}$ data, nominal Monte Carlo (MC) simulation (pre-fit), and post-fit Monte Carlo simulation comparisons in VA, $\left | t \right |$, Q$^{2}$, $p_{\pi}$, and cos($\theta_{\pi}$) for the signal region. The stacked histograms represent the true reaction types of the events. The smaller post-fit  $\chi^{2}_{\textup{stat.}}$ shows improved data and MC agreement in all five variables after the fit. The post-fit $\chi^{2}_{\textup{stat.}}$ for the VA distribution remains statistically incompatible to the data. The nominal MC was modified to reproduce the shapes seen in the data and used as a simulated data set. No significant bias in the total cross section extraction was observed (see text). Note that the fit does not consider these distributions, but only sees a single bin for the signal sample.}
\label{fig:nu_fitter_data_postfit_SIG}
\end{figure*}

\begin{figure*}[!htbp]
\centering
\includegraphics[width=0.9\textwidth]{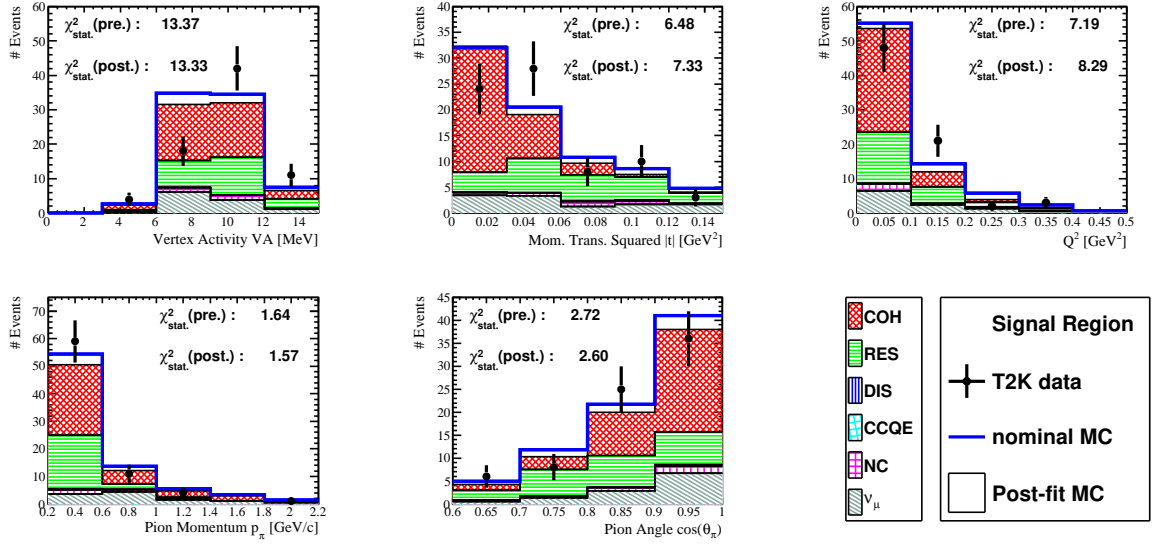}
\caption{$\bar{\nu}_{\mu}$ data, nominal Monte Carlo (MC) simulation (pre-fit), and post-fit Monte Carlo simulation comparisons in VA, $\left | t \right |$, Q$^{2}$, $p_{\pi}$, and cos($\theta_{\pi}$) for the signal region. The stacked histograms represent the true reaction types of the events. The post-fit $\chi^{2}_{\textup{stat.}}$ are not changed as much as in the $\nu_\mu$ case due to lower statistics. As with the $\nu_\mu$ case, only the VA distribution is statistically incompatible with the data. Note that the fit does not consider these distributions, but only sees a single bin for the signal sample.}
\label{fig:nubar_fitter_data_postfit_SIG}
\end{figure*}

To verify whether the MC model is suitable to describe the kinematics (and thus the selection) of the relevant events,
the data was compared with MC in distributions of $\left | t \right |$, VA, $Q^{2}$, $p_{\pi}$, and cos($\theta_{\pi}$). Note that the latter four variables are not used by the likelihood fit.
While there was mostly good agreement in the sample (see Fig.~\ref{fig:nu_fitter_data_postfit_SIG} and Fig.~\ref{fig:nubar_fitter_data_postfit_SIG}),
a disagreement in the shapes of the sideband samples was observed.
To evaluate what, if any, bias might be introduced due to the difference between the nominal MC and data in the sideband samples, several empirical tunings of the MC were made to obtain better agreement with the data. Among these, the following combination of changes in the nominal MC produced the best agreement with data:

\begin{enumerate}
    \item the VA of all background events was increased by 1 MeV, 
    \item additional VA, uniformly distributed between 0 and 100 MeV, was randomly added to 25\% of the interactions on neutron target,
    \item low-$Q^2$ ($< 0.7 ~\rm{GeV}^{2}$) CC-RES events were suppressed (using a MINERvA-inspired data driven suppression technique ~\cite{minervaLowq2}),
    \item the CC-DIS contribution was suppressed by 50\%,
    \item the CC-RES contribution was increased by 10\% and
    \item the normalisation of the signal COH events was increased by (18)19\% for (RHC)FHC mode, where the numbers were derived from the fit results discussed below.
\end{enumerate}

The need for a modified VA has also previously been seen at MINERvA~\cite{MINERvAPhysRevLett.113.261802} and the 2016 T2K CC-COH measurement~\cite{T2K:2016soz}. In addition to this, separate simulated data sets were created where the impacts of single parameter variations were studied.
None of the studies revealed any significant potential for bias in the cross section extraction, and differences between extracted and true cross section were always within the systematic uncertainties.

\begin{figure*}[!htbp]
\centering
\includegraphics[width=0.9\textwidth]{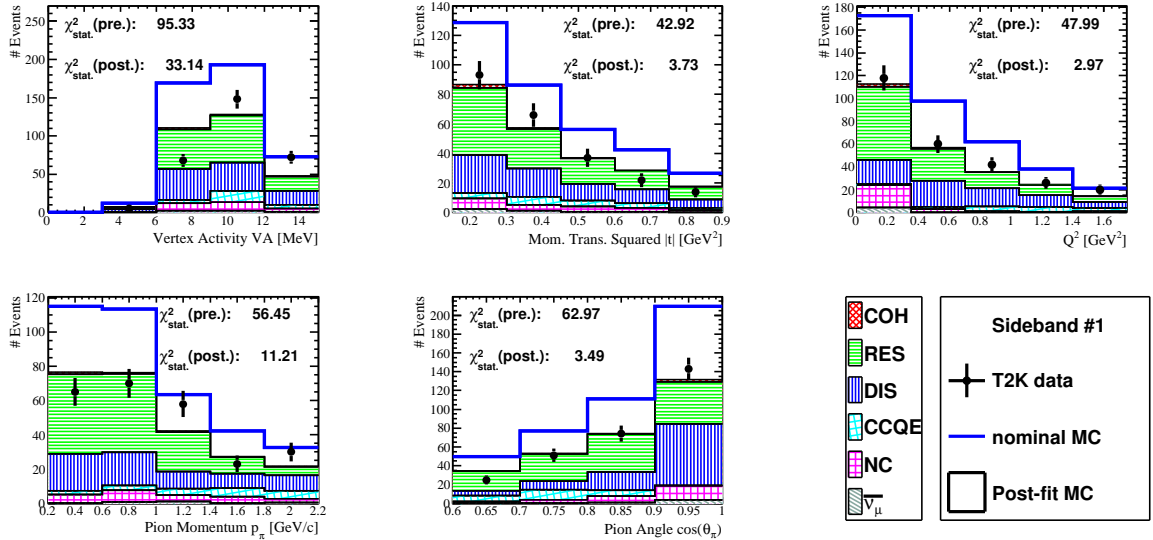}
\caption{$\nu_{\mu}$ data, nominal Monte Carlo (MC) simulation (pre-fit), and post-fit Monte Carlo simulation comparisons in VA, $\left | t \right |$, Q$^{2}$, $p_{\pi}$, and cos($\theta_{\pi}$) for the first sideband. The stacked histograms represent the true reaction types of the events. The post-fit $\chi^{2}_{\textup{stat.}}$ are improved in general as a result of the RES and DIS background events being re-weighted. The relatively larger VA and $p_{\pi}$ post-fit $\chi^{2}_{\textup{stat.}}$ indicate lack of degrees of freedom in these two spaces.}
\label{fig:nu_fitter_data_postfit_SB1}
\end{figure*}

\begin{figure*}[!htbp]
\centering
\includegraphics[width=0.9\textwidth]{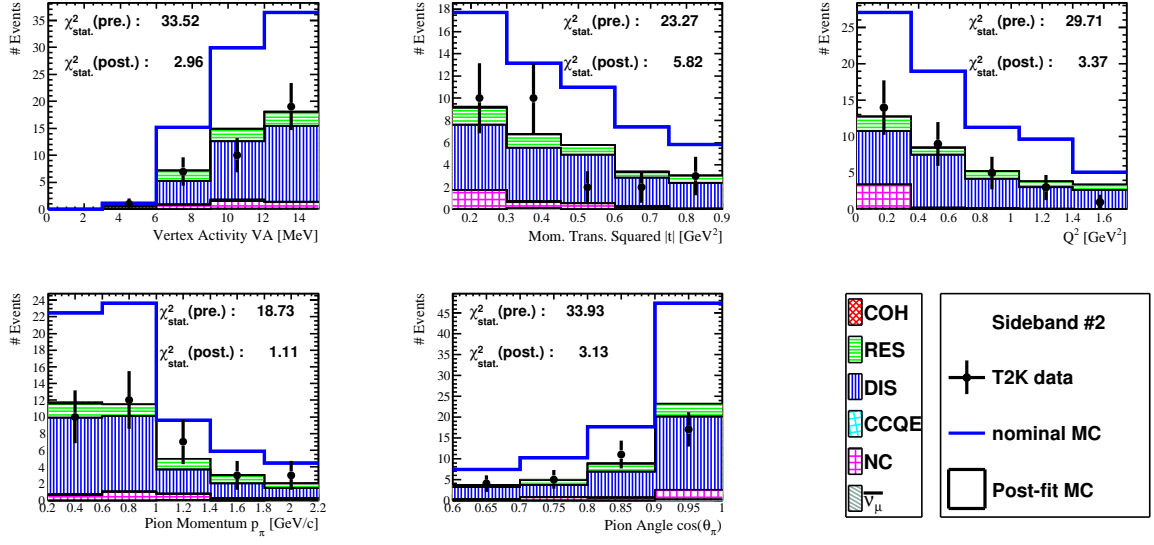}
\caption{$\nu_{\mu}$ data, nominal Monte Carlo (MC) simulation (pre-fit), and post-fit Monte Carlo simulation comparisons in VA, $\left | t \right |$, Q$^{2}$, $p_{\pi}$, and cos($\theta_{\pi}$) for the second sideband. The stacked histograms represent the true reaction types of the events. Since this sideband is dominated by the DIS background events, the post-fit $\chi^{2}_{\textup{stat.}}$ is much improved after the fitter as a result of the DIS background events being re-weighted.}
\label{fig:nu_fitter_data_postfit_SB2}
\end{figure*}

\begin{figure*}[!htbp]
\centering
\includegraphics[width=0.9\textwidth]{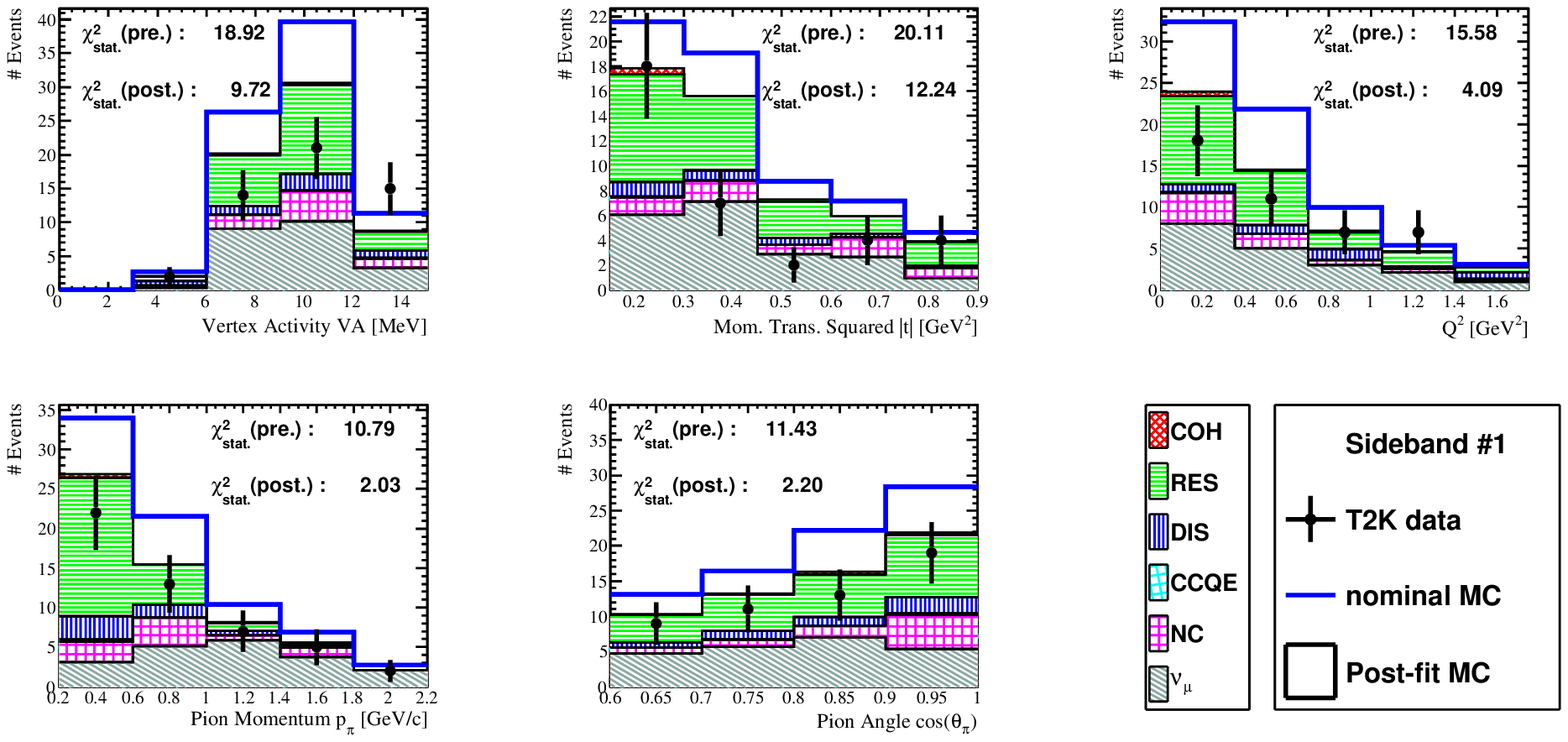}
\caption{$\bar{\nu}_{\mu}$ data, nominal Monte Carlo (MC) simulation (pre-fit), and post-fit Monte Carlo simulation comparisons in VA, $\left | t \right |$, Q$^{2}$, $p_{\pi}$, and cos($\theta_{\pi}$) for the first sideband. The stacked histograms represent the true reaction types of the events. The post-fit $\chi^{2}_{\textup{stat.}}$ are improved in general as a result of the RES background events being re-weighted; however, the shape differences between the data and the MC in the VA and $\left | t \right |$ cannot be simply resolved by simple re-scale of events.}
\label{fig:nubar_fitter_data_postfit_SB1}
\end{figure*}

An additional simulated data study was motivated by the report of a suppression of CC-$1\pi$ resonant events at low $Q^2$ from the MINERvA experiment\cite{minervaLowq2}.
This effect has not been observed or excluded by T2K, which has different neutrino beam energies and nominal MC models, but the potential effect of such an observation in T2K on the cross section result was studied nonetheless.
Analysis of a simulated data set with an artificially suppressed CC-RES cross section at low $Q^2$ resulted in a noticeable bias in the extracted cross section, which was not covered by the model uncertainties considered in this analysis.
Since it is unclear whether this suppression should be expected in the T2K data, the bias seen in this study is covered with an additional systematic uncertainty on the extracted (anti)neutrino CC-COH cross section of 16\% (24\%).
The size of this uncertainty was chosen so that the
one standard deviation of the total systematic uncertainty, including that attributed to the low $Q^2$ CC-RES suppression, covers the bias observed in this simulated data study.

The $\nu_{\mu}$ analysis employs three bins in reconstructed parameter space (one bin from the signal region and two bins from the two sidebands). The $\chi^{2}_{\textup{post-fit.}}$ (9.44) improved significantly from the $\chi^{2}_{\textup{pre-fit.}}$  (82.53).
The $\chi^{2}_{\textup{pre-fit.}}$ is not small due to large difference between data and MC in one of the sidebands as a result of over-prediction of the DIS background events.
The $\bar{\nu}_{\mu}$ analysis employs two bins in reconstructed space (one bin each from the signal region and the sideband).
The $\chi^{2}_{\textup{post-fit.}}$ (4.62) improved from the $\chi^{2}_{\textup{pre-fit.}}$  (9.79).
These values are within the range of results from simulated data studies that showed acceptable levels of bias in the cross section results. 

As shown in Fig.~\ref{fig:nu_fitter_data_postfit_SIG},~\ref{fig:nu_fitter_data_postfit_SB1}, and~\ref{fig:nu_fitter_data_postfit_SB2}, the $\chi^{2}_{\textup{stat}}(\textup{post-fit})$ are reduced from the $\chi^{2}_{\textup{stat}}(\textup{pre-fit})$ in most kinematic distributions of the $\nu_{\mu}$ data. The improvements are the result of the RES and DIS background events being re-weighted. The improvements are especially obvious for the two sidebands, which are mostly consist of the RES and DIS background events. The relatively larger VA and $p_{\pi}$ post-fit $\chi^{2}_{\textup{stat.}}$ indicate lack of degrees of freedom in these two spaces. Similar conclusions can be drawn from the $\bar{\nu}_{\mu}$ data shown in Fig.~\ref{fig:nubar_fitter_data_postfit_SIG} and~\ref{fig:nubar_fitter_data_postfit_SB1}. However, the improvements in the post-fit $\chi^{2}_{\textup{stat.}}$ are less significant since the bin-to-bin uncertainties are dominated by low statistics.
The $\chi^{2}_{\textup{stat.}}$ shown in these figures are only used to indicate whether there has been an improvement in the agreement between data and model after the fit has been performed. 
Since the fit only uses a single bin for the signal region in each of the two sidebands,
any improvement in the shapes of these distributions comes mostly from the relative normalisation of the background DIS and RES interaction modes. 

The fitted flux-averaged, phase space-constrained, charged current coherent cross sections per atom of the FGD FV are shown in Table~\ref{tab:results}. These results are based on an event sample of (80) 272 events with a predicted background component of (46) 159 events in the (RHC) FHC beam mode. The table includes the result on the carbon cross section derived using the $A^{1/3}$ scaling and model predictions for comparison.
The measured cross sections on the FGD target material agree with the NEUT Berger-Sehgal prediction slightly better than with the GENIE Rein-Sehgal model, but they are both covered by the measurement uncertainties. 

\begin{table*}[!htbp]
    \centering
    \caption{Summary of flux-averaged, phase-space-constrained, charged current, coherent cross section results and model predictions. Expressed in units of $10^{-40}$~$\mbox{cm}^{2}$/nucleus with statistical and systematic uncertainties, and the additional uncertainty added to cover the low Q$^{2}$ CC-RES suppression bias. The reported measurement on carbon uses the $F_{1/3}(A)$ scaling.
    The model predictions for carbon do not use a scaling function.  No prediction in RHC mode is given using the GENIE RS (2007) model as  Monte Carlo simulation data sets using the GENIE RS model had not been generated for the RHC beam mode at the time of this analysis. Note that the cross section prediction from the NEUT BS (2009) model for the FHC mode is different from the prediction for the RHC mode as the flux of neutrinos and antineutrinos are different.  }
    \begin{tabular}{l@{\quad}ccc} \hline
         & T2K (2022) & NEUT BS (2009) & GENIE RS (2007) \\ \hline
        $\sigma_{\nu_\mu,\text{FGD}}$       & $3.00 \pm 0.37 \pm 0.31 \pm 0.49$   & 2.77 & 3.28 \\
        $\sigma_{\nu_\mu,\text{C},1/3}$ & $2.98 \pm 0.37 \pm 0.31 \pm 0.49$       & 2.57 & 3.09 \\
        $\sigma_{\bar{\nu}_\mu,\text{FGD}}$ & $3.07 \pm 0.71 \pm 0.39 \pm 0.75$   & 2.87 &  /   \\
        $\sigma_{\bar{\nu}_\mu,\text{C},1/3}$ & $3.05 \pm 0.71 \pm 0.39 \pm 0.74$ & 2.78 &  /  \\ \hline
    \end{tabular}
    \label{tab:results}
\end{table*}

The sources of uncertainties are summarised in Table~\ref{tab:systematic}. The systematic uncertainty is further broken down into three components: the flux uncertainties, the cross section and final state interaction (FSI) related uncertainties, and the detector response related uncertainties. The total systematic uncertainty quoted in the table reflects the correlations between the three components. An additional uncertainty is added to cover the low Q$^{2}$ CC-RES suppression bias as described previously. The main contributions to the uncertainty for both neutrino and antineutrino results come from statistical uncertainties and the additional low Q$^{2}$ CC-RES suppression uncertainty.

\begin{table*}[!htbp]
    \centering
    \caption{Statistical uncertainty and breakdown of the sources of systematic uncertainties. The largest contribution of uncertainty comes from the bias in the extracted cross section when the low Q$^{2}$ CC-RES events are suppressed. Note the total systematic uncertainty is not exactly equal to the quadratic sum of the components due to correlation between the sources.}
    \begin{tabular}{cc@{\quad}c} \hline
         Sources of Uncertainties & $\nu_{\mu}$ CC-COH ($\times$ 10$^{-40}$ cm$^{2})$ & $\bar{\nu}_{\mu}$ CC-COH ($\times$ 10$^{-40}$ cm$^{2}$)\\
        \hline
            Flux & 0.14 & 0.24\\
            cross section and FSI & 0.22 & 0.34\\   
            Detector Responses & 0.24 & 0.42\\
        \hline
            Total Systematic Uncertainty &  0.31 & 0.39 \\
            Low Q$^{2}$ CC-RES Suppression Related & 0.49 & 0.75\\
            Statistical & 0.37 & 0.71\\ \hline
    \label{tab:systematic}
    \end{tabular}
\end{table*}

\begin{figure*}[!htbp]
\centering
\includegraphics[width=0.45\textwidth]{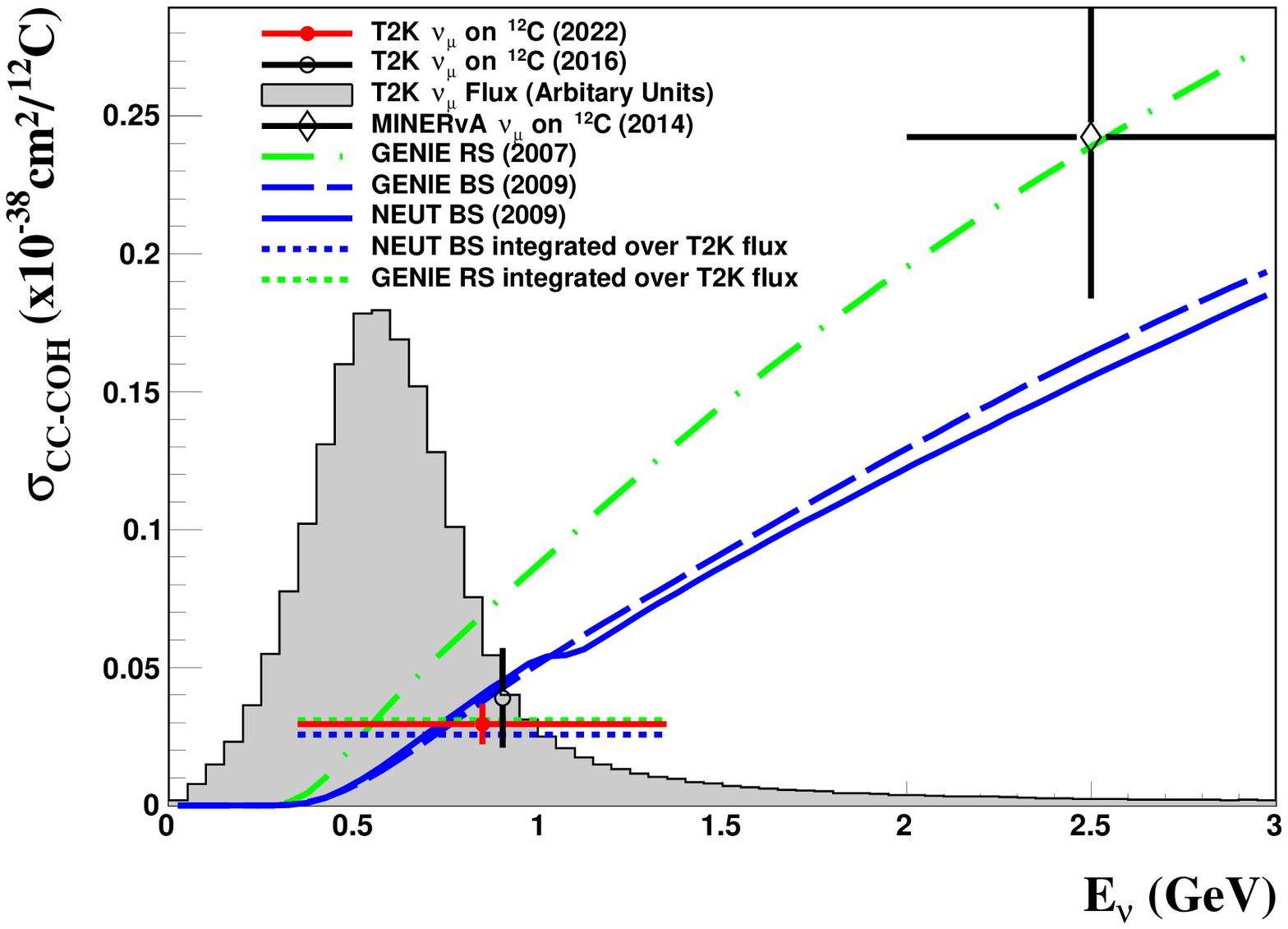}
\includegraphics[width=0.45\textwidth]{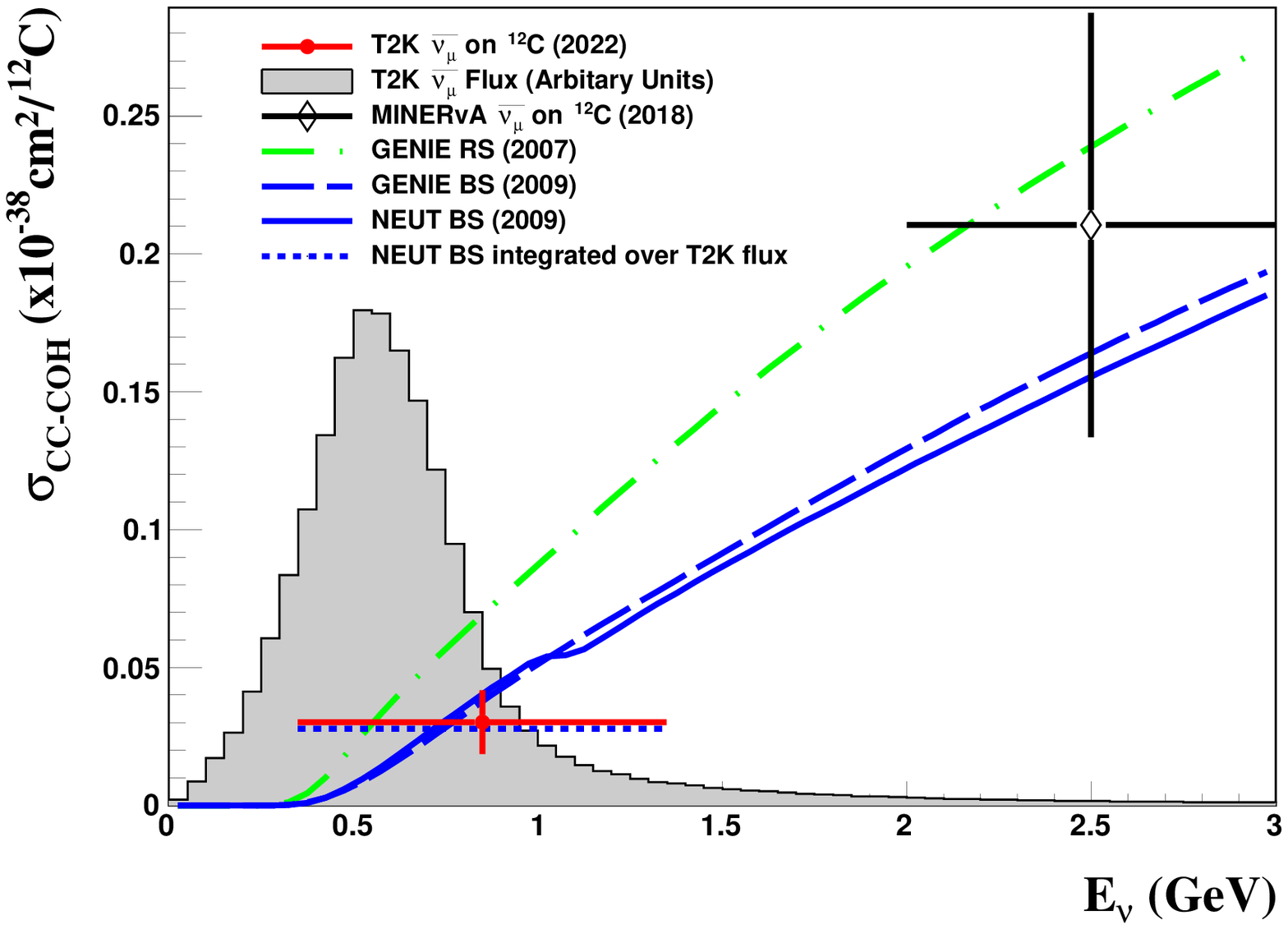}
\caption{The T2K $\nu_{\mu}$ (left) and $\bar{\nu}_{\mu}$ (right)  CC-COH cross section measurement on C assuming $F(A)= A^{1/3}$. The measurement uncertainty shown is the quadratic sum of the statistical and systematic components. The $x$-axis error bar covers one standard deviation of the T2K flux around the mean (anti)neutrino energy of $(0.85)0.85$~GeV. The 2016 $\nu_{\mu}$ T2K result (same E$_{\nu}$ range, center point offset slightly for clarity of the figure)~\cite{T2K:2016soz}, and the MINERvA result~\cite{MINERvA:2017ipy} (for $E_{\nu}$ between 2--3 GeV) are also shown for comparison. Note the phase space between the different measurements are not exactly the same.}
\label{fig:xsec_c}
\end{figure*}

The cross section results are shown in Fig.~\ref{fig:xsec_c}.
The cross section reported here is energy-averaged over the incoming neutrino flux and restricted to a specific region of the parameter space of produced muon and pion kinematics. As such, it cannot be directly compared to a theoretical model providing the cross section as a function of neutrino energy.
A valid comparison requires the theoretical cross section to be integrated over the T2K flux and phase space restrictions applied. Horizontal lines in Fig.~\ref{fig:xsec_c} show model predictions after this procedure has been applied. 

To enable a quick comparison between results of different experiments with different neutrino energy distributions, the mean neutrino energy is used as the $x$-position of the data points and the standard deviation of neutrino energies as the error bars in the $x$-direction.

\section{Conclusion}

The T2K $\nu_{\mu}$ CC-COH and $\bar{\nu}_{\mu}$ CC-COH cross-sections on $^{12}\textup{C}$ are
$(2.98 \pm 0.37 \textup{(stat.)} \pm 0.31\textup{(syst.)}\,\substack{ +0.49 \\ -0.00 }\,\mathrm{ (Q^2\,model)})  \times 10^{-40}~\mathrm{cm}^{2}$
and 
$(3.05 \pm 0.71 \textup{(stat.)} \pm 0.39 \textup{(syst.)}\,\substack{ +0.74 \\ -0.00 }\,\mathrm{(Q^2\,model)}) \times 10^{-40}~\mathrm{cm}^{2}$, 
assuming an $A$-scaling of $A^{1/3}$. The flux-averaged cross-sections are measured in a reduced final state particle kinematic phase space for which $p_{\mu, \pi}$ $>$ 0.2~GeV, $\textup{cos}(\theta_{\mu})$ $>$ 0.8, and $\textup{cos}(\theta_{\pi})$ $>$ 0.6. The uncertainty labelled as ``Q$^2$ model'' corresponds to the potential bias caused by the low Q$^{2}$ CC-RES suppression study. 
This study presents the first measurement of the $\bar{\nu}_{\mu}$ CC-COH cross-section at a mean neutrino energy less than 1~GeV. In addition,
the $\nu_{\mu}$ CC-COH measurement is consistent with the previous 2016 T2K result but with the fractional total uncertainty reduced from 46\% to 23\%. It is notable that the measured neutrino and antineutrino coherent pion production cross-sections are themselves consistent, as expected from theory.
Both the NEUT Berger-Sehgal and the GENIE Rein-Sehgal model predictions are compatible with the data within the measurement uncertainties.

A data release summarising these results is available from the T2K public results site\cite{T2KCoherentDR:2023}.

\begin{acknowledgments}
We thank the J-PARC staff for superb accelerator performance. We thank the CERN NA61/SHINE Collaboration for providing valuable particle production data. We acknowledge the support of MEXT,   JSPS KAKENHI (JP16H06288, JP18K03682, JP18H03701, JP18H05537, JP19J01119, JP19J22440, JP19J22258, JP20H00162, JP20H00149, JP20J20304) and bilateral programs(JPJSBP120204806, JPJSBP120209601), Japan; NSERC, the NRC, and CFI, Canada; the CEA and CNRS/IN2P3, France; the DFG (RO 3625/2), Germany; the INFN, Italy; the Ministry of Education and Science(DIR/WK/2017/05) and the National Science Centre (UMO-2018/30/E/ST2/00441 ), Poland; the RSF19-12-00325, RSF22-12-00358, Russia; MICINN (SEV-2016-0588, PID2019-107564GB-I00, PGC2018-099388-BI00, PID2020-114687GB-I00) Government of Andalucia (FQM160, SOMM17/6105/UGR) and the University of Tokyo ICRR's Inter-University Research Program FY2023 Ref. J1, and ERDF funds and CERCA program, Spain; the SNSF and SERI (200021\_185012, 200020\_188533, 20FL21\_186178I), Switzerland; the STFC and UKRI, UK; and the DOE, USA. We also thank CERN for the UA1/NOMAD magnet, DESY for the HERA-B magnet mover system, the BC DRI Group, Prairie DRI Group, ACENET, SciNet, and CalculQuebec consortia in the Digital Research Alliance of Canada, and GridPP in the United Kingdom, and the CNRS/IN2P3 Computing Center in France. In addition, the participation of individual researchers and institutions has been further supported by funds from the ERC (FP7), “la Caixa” Foundation (ID 100010434, fellowship code LCF/BQ/IN17/11620050), the European Union’s Horizon 2020 Research and Innovation Programme under the Marie Sklodowska-Curie grant agreement numbers 713673 and 754496, and H2020 grant numbers RISE-GA822070-JENNIFER2 2020 and RISE-GA872549-SK2HK; the JSPS, Japan; the Royal Society, UK; French ANR grant number ANR-19-CE31-0001; the SNF Eccellenza grant number PCEFP2\_203261; and the DOE Early Career programme, USA. For the purposes of open access, the authors have applied a Creative Commons Attribution licence to any Author Accepted Manuscript version arising.
\end{acknowledgments}

\bibliography{CCCoherent}

\end{document}